\documentclass[12pt]{article}

\usepackage{ifpdf}
\ifpdf
\usepackage{graphicx,color}
\usepackage{hyperref}
\else
\fi
\usepackage{amssymb,amsfonts,amsmath,cancel,cite,multirow}
\usepackage[capitalise]{cleveref}
\usepackage{slashed}

\newcommand{\eqs}[1]{\begin{equation}\begin{split} #1 \end{split}\end{equation}}

\begin{document}

\begin{titlepage}

\begin{flushright}
	CTPU-PTC-19-04
\end{flushright}

\vskip 1.35cm
\begin{center}

{\large
\textbf{
On Scalaron Decay via the Trace of Energy-Momentum Tensor}}
\vskip 1.2cm

Ayuki Kamada$^{a}$
\vskip 0.4cm

\textit{$^a$
Center for Theoretical Physics of the Universe,
Institute for Basic Science (IBS), Daejeon 34126, Korea
}

\vskip 1.5cm

\begin{abstract}
In some inflation scenarios such as $R^{2}$ inflation, a gravitational scalar degrees of freedom called scalaron is identified as inflaton.
Scalaron linearly couples to matter via the trace of energy-momentum tensor.
We study scenarios with a sequestered matter sector, where the trace of energy-momentum tensor predominantly determines the scalaron coupling to matter.
In a sequestered setup, heavy degrees of freedom are expected to decouple from low-energy dynamics.
On the other hand, it is non-trivial to see the decoupling since scalaron couples to a mass term of heavy degrees of freedom.
Actually, when heavy degrees of freedom carry some gauge charge, the amplitude of scalaron decay to two gauge bosons does not vanish in the heavy mass limit.
Here a {\it quantum} contribution to the trace of energy-momentum tensor plays an essential role.
This quantum contribution is known as trace anomaly or Weyl anomaly.
The trace anomaly contribution from heavy degrees of freedom cancels with the contribution from the {\it classical} scalaron coupling to a mass term of heavy degrees of freedom.
We see how trace anomaly appears both in the Fujikawa method and in dimensional renormalization.
In dimensional renormalization, one can evaluate the scalaron decay amplitude in principle at all orders, while it is unclear how to process it beyond the one-loop level in the Fujikawa method.
We consider scalaron decay to two gauge bosons via the trace of energy-momentum tensor in quantum electrodynamics with scalars and fermions.
We evaluate the decay amplitude at the leading order to demonstrate the decoupling of heavy degrees of freedom.
\end{abstract}

\end{center}
\end{titlepage}

\section{Introduction}
Inflation is a cosmological paradigm that solves issues of big bang cosmology, such as the horizon, flatness, and monopole problems~\cite{Brout:1977ix, Starobinsky:1980te, Kazanas:1980tx, Sato:1980yn, Guth:1980zm, Linde:1981mu, Albrecht:1982wi, Linde:1983gd}.
It also provides an almost scale-invariant density contrast over homogeneous and isotropic background~\cite{Mukhanov:1981xt, Mukhanov:1982nu, Hawking:1982cz, Guth:1982ec, Starobinsky:1982ee, Bardeen:1983qw, Mukhanov:1985rz}.
The inflation paradigm has been strongly supported by the deviation of the scalar spectral index from unity observed in cosmic microwave background anisotropies~\cite{Akrami:2018odb}.
Among various inflation models~\cite{Martin:2013tda}, $R^{2}$ inflation ($R$: Ricci scalar)~\cite{Starobinsky:1980te, Barrow:1983rx, Whitt:1984pd, Vilenkin:1985md, Mijic:1986iv, Barrow:1988xh} is a good benchmark.
Its plateau potential predicts a tensor-to-scalar ratio sufficiently small to be consistent with the Planck data~\cite{Akrami:2018odb} but within a reach of future searches of cosmic microwave background $B$-mode anisotropies~\cite{Matsumura:2016sri, Delabrouille:2017rct, Abazajian:2016yjj}.

Identifying a reheating temperature $T_{R}$ in $R^{2}$ inflation is important for theoretical prediction of the scalar spectral index and tensor-to-scalar ratio~\cite{Bezrukov:2011gp}.
It also plays an important role in production mechanisms of dark matter and baryon asymmetry~\cite{Gorbunov:2010bn, Gorbunov:2012ij}.
For example, $T_{R} \gtrsim 10^{9} \, {\rm GeV}$ (e.g., Refs~\cite{Buchmuller:2002rq, Giudice:2003jh, Buchmuller:2005eh, Davidson:2008bu}) is required for thermal leptogenesis~\cite{Fukugita:1986hr} to work.
Furthermore it is imprinted in the primordial gravitational wave spectrum when the energy density of Universe is transferred from oscillating inflaton to radiation~\cite{Gorbunov:2012ns}.
Such an imprint could be seen in ultimate gravitational wave experiments~\cite{Kawamura:2006up}.

In $f (R)$ gravity including $R^{2}$ inflation, a gravitational scalar degrees of freedom called scalaron is identified as inflaton.
To determine the reheating temperature, we need to study scalaron coupling to matter.
$f (R)$ gravity generically can be rewritten as a scalar-tensor theory through a Weyl transformation (local rescaling of the metric and fields) that is a function solely of scalaron~\cite{Whitt:1984pd, Jakubiec:1988ef}.
This Weyl transformation manifests scalaron coupling to the trace of matter energy-momentum tensor in the {\it scalaron frame}~\cite{Faulkner:2006ub}.%
\footnote{
Note that the Weyl transformation consists solely of scalaron.
Therefore this is not the Einstein frame since scalar fields in a matter section (not scalaron) can still have a non-minimal coupling to the Ricci scalar.
}
The trace of energy-momentum tensor predominantly determines the scalaron coupling to matter.
Similar situations can also be seen in a broader class of inflation models based on a scalar-tensor theory.
One example is $f(\sigma) R$ gravity (let us also refer to a scalar field $\sigma$ as scalaron)~\cite{Accetta:1985du, La:1989za, Futamase:1987ua, Salopek:1988qh, Fakir:1990eg}.
In the scalaron frame, again, scalaron manifestly couples to the trace of matter energy-momentum tensor.
The trace of energy-momentum tensor can predominantly determine scalaron coupling to matter, when scalaron direct coupling to matter in the Jordan frame is suppressed for some reason.
In this paper, we consider such scenarios where a matter sector communicates with the scalaron sector only gravitatioanlly in the Jordan frame.

Scalaron decay%
\footnote{
In this paper, we consider {\it perturbative} scalaron decay.
We assume that non-perturbative effects associated with non-zero field values of scalaron and matter scalars are negligible.
This could be true since decay proceeds only gravitationally and occurs long after inflation.}
 is dominated by decay channels to two scalars if their non-minimal coupling to Ricci curvature deviates from the conformal coupling.
With the conformally coupled scalars, loop-induced decay to two gauge bosons becomes relevant.
The decay amplitude is proportional to the $\beta$ function of the corresponding gauge coupling.
Ref.~\cite{Gorbunov:2012ns} uses the $\beta$ function at the energy scale of the scalaron mass ($\simeq 3 \times 10^{13} \, {\rm GeV}$ for the $R^{2}$ inflation model), which virtually counts {\it light} degrees of freedom.
Refs.~\cite{Kannike:2015apa, Choi:2019osi}, which study inflaton decay in $f(\sigma) R$ gravity, virtually counts {\it light} degrees of freedom.

On the other hand, it is non-trivial if heavy degrees of freedom do not contribute to the scalaron decay.
In the scalaron frame, scalaron couples to matter via mass terms.
Loop-induced decay to two gauge bosons does not vanish in the heavy mass limit.
It leaves scalaron coupling to gauge bosons for low-energy effective theory.%
\footnote{
This is the case for Higgs~\cite{Ellis:1975ap, Shifman:1979eb} or axion~\cite{Peccei:1977hh, Peccei:1977ur, Weinberg:1977ma, Wilczek:1977pj} (see also Refs.~\cite{Kim:1979if, Shifman:1979if} and \cite{Zhitnitsky:1980tq, Dine:1981rt} for popular ultraviolet realizations).
One famous example is coupling of Higgs~\cite{Ellis:1975ap, Shifman:1979eb} or axion~\cite{Peccei:1977hh, Peccei:1977ur, Weinberg:1977ma, Wilczek:1977pj} (see also Refs.~\cite{Kim:1979if, Shifman:1979if} and \cite{Zhitnitsky:1980tq, Dine:1981rt} for popular ultraviolet realizations) to light gauge bosons such as photon or gluon in low-energy effective theory.
With this observation, Ref.~\cite{Takeda:2014qma} argues that one should count {\it heavy} degrees of freedom as well as {\it light} degrees of freedom for the $\beta$ function.
This result is taken from Ref.~\cite{Watanabe:2010vy}, which studies inflaton decay in $f(\sigma) R$ gravity.
A similar calculation on scalaron coupling to the standard model particles has been made in Ref.~\cite{Katsuragawa:2016yir}.
Their stance on the frame equivalence is different from the present study.
}
Meanwhile the decoupling of heavy degrees of freedom may be apparent in the Jordan frame, where scalaron does not have any direct coupling to matter.
Matter fields decouple in the heavy mass limit without leaving any non-decoupling effects for low-energy effective theory.
This raises an issue on the ``frame equivalence'' (see also Ref.~\cite{Falls:2018olk} for a related discussion).

What plays an essential role is a {\it quantum} contribution to the trace of energy-momentum tensor, known as Weyl anomaly or trace anomaly.%
\footnote{
The trace of energy-momentum tensor and trace anomaly are often not distinguished.
In this paper we use the former to refer to the whole (classical + quantum) contribution, while we use the latter to refer to only a quantum contribution.
}
Trace anomaly is intensively investigated both in the flat spacetime~\cite{Callan:1970ze, Coleman:1970je, Freedman:1974gs, Freedman:1974ze, Collins:1976vm, Nielsen:1977sy, Adler:1976zt, Collins:1976yq, Brown:1979pq, Brown:1980qq, Hathrell:1981zb, Hathrell:1981gz} and in a curved spacetime~\cite{Capper:1974ed, Deser:1974cz, Capper:1974ic, Capper:1973mv, Dowker:1976zf, Brown:1976wc, Christensen:1977jc, Brown:1977pq, Duff:1977ay, Bunch:1978yq} (see also Ref.~\cite{Duff:1993wm} for a review).
The trace anomaly contribution from heavy degrees of freedom cancels with the contribution from the {\it classical} scalaron coupling to a mass term of heavy degrees of freedom.
Because of the cancellation between classical and quantum contributions, the scalaron coupling to matter via the trace of energy-momentum tensor is ultraviolet insensitive.%
\footnote{
This is analogous to an anomaly mediation contribution to a sparticle mass in supersymmetric theories~\cite{Randall:1998uk, Giudice:1998xp}, which boasts its ultraviolet insensitivity.
A quantum contribution to a gaugino mass from heavy degrees of freedom cancels with a classical contribution from a coupling of a compensator field to a mass term of heavy degrees of freedom.
Indeed superconformal anomaly is correctly taken into account in supersymmetric inflation setups~\cite{Endo:2007ih, Endo:2007sz, Terada:2014uia}.
}

This paper is organized as follows.
In the next section we describe scenarios with a sequestered matter sector, where scalaron couples to matter predominantly via the trace of energy-momentum tensor.
We demonstrate how the trace of energy-momentum tensor receives a quantum contribution, by employing the Fujikawa method~\cite{Fujikawa:1980vr, Fujikawa:1980rc, Fujikawa:1993xv} (see also Ref.~\cite{Fujikawa:2004cx} for a comprehensive summary).
The Fujikawa method is illustrating trace anomaly, but not convenient in practical calculations such as perturbative renormalization.
Instead, in \cref{sec:flattraceT}, we use dimensional renormalization, i.e., the minimal subtraction (MS) or modified minimal subtraction ($\overline{\rm MS}$) scheme~\cite{Bollini:1972ui, Ashmore:1972uj, tHooft:1972tcz}, where we can compute perturbative renormalization in principle at all orders.
\footnote{
Ref.~\cite{Choi:2019osi} sketches the derivation of trace anomaly at the one-loop order in Wilsonian renormalization.
} 
We see how trace anomaly appears in dimensional renormalization.
Furthermore, we compute the leading amplitude of scalaron decay into two gauge boson in quantum electrodynamics (QED) with scalars and fermions.
We see that heavy degrees of freedom do not contribute to the amiplitude.
\cref{sec:concl} is devoted to a summary and further remarks.
We use a notation of Ref.~\cite{Kolb:1990vq}, where the four-dimension metric has the signature of $(+ , - , - , -)$.

%%%%%%%%%%%%%%%%%%%%%%%%%%%%%%%%%%%%%%%%%%%%%%%%%%%%%%%%%%%%%%%%%%%%%%%%%%%%%%%%%
\section{Gravitational coupling of scalaron to matter \label{sec:sigtomat}}
We consider a class of inflation models where a scalaron sector communicates with a matter sector only gravitationally as
\eqs{
\label{eq:actionp}
S_{\rm grav} \left[ g'_{\mu \nu}, \sigma' \right] + S_{\rm mat} \left[ \{ \phi'_{i} \}, g'_{\mu \nu}; \{ \lambda_{a} \} \right] \,,
}
in the Jordan frame.
$g_{\mu \nu}$ is the metric.
$\{ \phi_{i} \}$ and $\{ \lambda_{a} \}$ collectively denote matter fields and parameters, respectively.
Note that scalaron in $f(R)$ gravity is not manifest in the Jordan frame.
For example, in the $R^{2}$ inflation model,
\eqs{
S_{\rm grav} = - \frac{M_{\rm pl}^{2}}{2} \int d^{4} x \sqrt{- g'} \left( R' - \frac{R'^{2}}{6 \mu^{2}} \right) \,,
}
with the reduced Planck mass $M_{\rm pl} \simeq 2.435 \times 10^{18} \, {\rm GeV}$ and a mass parameter $\mu$.
We assume that the matter sector is minimally coupled to gravity, while maintaining renormalizability up to graviton loops that are suppressed by $1 / M_{\rm pl}^{2}$.%
\footnote{This does not mean the matter sector consists solely of a finite number of renormalizable terms.
Non-renormalizable terms are allowed when an infinite number of non-renormalizable terms are introduced for renormalization in the usual sense of effective field theory.}
In particular we require renormalizablity of energy-momentum tensor that is defined as a linear response of the matter action to the metric.
For example, QED with a scalar $\phi$ is described by
\eqs{
S_{\rm mat} =& \int d^{4} x \sqrt{- g'} \left( - \frac{1}{4} g'^{\mu \lambda} g'^{\nu \kappa} F'_{\mu \nu} F'_{\lambda \kappa} +  g'^{\mu \nu} D'_{\mu} \phi'^{*} D'_{\nu} \phi' + \xi_{\rm grav} R' |\phi'|^{2} - m_{s}^{2} |\phi'|^{2} - \frac{1}{4} \lambda |\phi'|^{4} \right) \\ 
& + S_{\rm fix} \,,
}
with $D_{\mu}$ being the gauge and diffeomorphism covariant derivative and $F_{\mu \nu}$ being the field strength of $A_{\mu}$.
$m_{s}$ is a scalar mass and $\lambda$ is a quartic coupling. 
A non-minimal coupling $\xi_{\rm grav}$, which provides an improvement term of energy-momentum tensor~\cite{Callan:1970ze, Coleman:1970je}, should be kept to maintain renormalizability of energy-momentum tensor.
We devote \cref{sec:gaugefix} to the gauge fixing term $S_{\rm fix}$, whose contribution to the energy-momentum tensor can be omitted for physical states.

The scalaron + gravity sector turns into the Einstein-Hilbert action + scalaron action via the Weyl transformation of
\eqs{
\label{eq:gravresponse}
g'_{\mu \nu} = e^{2 \omega (\sigma)}  g_{\mu \nu} \,.
}
The action in the scalaron frame is $S_{\rm grav} = S_{\text{E-H}} + S_{\sigma}$ where
\eqs{
& S_{\text{E-H}} = - \frac{M_{\rm pl}^{2}}{2} \int d^{4} x \sqrt{- g} R \,, \\
& S_{\sigma} = \int d^{4} x \sqrt{- g} \left( \frac{1}{2} g^{\mu \nu} \nabla_{\mu} \sigma \nabla_{\nu} \sigma - V(\sigma) \right) \,,
}
with $\nabla_{\mu}$ being the diffeomorphism covariant derivative.

For example, in the $R^{2}$ inflation model,
\eqs{
\omega = - \frac{1}{\sqrt{6}} \frac{\sigma}{M_{\rm pl}}
}
and
\eqs{
V(\sigma) = \frac{3}{4} \mu^{2} M_{\rm pl}^{2} \left[ 1 - \exp \left( - \sqrt{\frac{2}{3}} \frac{\sigma}{M_{\rm pl}} \right) \right]^{2} \,.
}

The matter fields transform under the Weyl transformation as
\eqs{
\phi'_{i} = e^{- d_{i} \omega (\sigma)}  \phi_{i} \,,
}
with $d_{i}$ denoting the Weyl weight of the field $\phi_{i}$.
The linear variation of the matter action is responsible for the leading coupling of scalaron to matter:
\eqs{
\label{eq:matresponse}
S_{\rm mat} \left[ \{ \phi'_{i} \}, g'_{\mu \nu}; \{ \lambda_{a} \} \right] \simeq& S_{\rm mat} \left[ \{ \phi_{i} \}, g_{\mu \nu}; \{ \lambda_{a} \} \right] - \int d^{4} x \sqrt{-g} \omega(\sigma) A_{\rm lin} \left( \{ \phi_{i} \}, g_{\mu \nu}; \{ \lambda_{a} \} \right) \,.
}
When we treat fields as classical objects, it is given by
\eqs{
A^{\rm class}_{\rm lin} = - \sum_{i} d_{i} ({\rm e.o.m.})_{i} + \left( g_{\mu \nu} T^{\mu \nu} \left( \{ \phi_{i} \}, g_{\mu \nu}; \{ \lambda_{a} \} \right) \right)_{\rm class} \,,
}
and
\eqs{
({\rm e.o.m.})_{i} = - \phi_{i} \frac{1}{\sqrt{- g}} \frac{\delta S_{\rm mat} \left[ \{ \phi_{i} \}, g_{\mu \nu}; \{ \lambda_{a} \} \right] }{\delta \phi_{i}} \,.
}
The second term of $A^{\rm class}_{\rm lin}$ is the {\it classical} trace of energy-momentum tensor, in which we treat fields as classical objects.
We define energy-momentum tensor by a functional derivative of
\eqs{
T^{\mu \nu} = - \frac{2}{\sqrt{-g}} \frac{\delta S_{\rm mat} \left[ \{ \phi_{i} \}, g_{\mu \nu}; \{ \lambda_{a} \} \right]}{\delta g_{\mu \nu}} \,.
}
For example, in scalar QED,
\eqs{
\label{eq:scalarcurvedTmunu}
T_{\mu \nu} =& - g^{\lambda \kappa} F_{\mu \lambda} F_{\nu \kappa} +  2 D_{\mu} \phi^{*} D_{\nu} \phi + 2 \xi_{\rm grav} R_{\mu \nu} |\phi|^{2} - 2 \xi_{\rm grav} \left( \nabla_{\mu} \nabla_{\nu} - g_{\mu \nu} g^{\lambda \kappa} \nabla_{\lambda} \nabla_{\kappa} \right) |\phi|^{2} \\
& - g_{\mu \nu} \left( - \frac{1}{4} g^{\lambda \rho} g^{\kappa \sigma} F_{\lambda \kappa} F_{\rho \sigma} + g^{\lambda \kappa} D_{\lambda} \phi^{*} D_{\kappa} \phi + \xi_{\rm grav} R |\phi|^{2} - m^{2} |\phi|^{2} - \frac{1}{4} \lambda |\phi|^{4} \right) \,.
}

When we treat fields as quantum operators, the linear variation $A_{\rm lin}$ receives an additional contribution $A_{\rm anom}$.
To see it, let us take a path integral formalism with path integral measure of ${\cal D} \{ \phi'_{i} \} [g'_{\mu \nu}]$.
Note that the path integral measure depends on the metric such that the path integral is diffeomorphism invariant~\cite{Fujikawa:2004cx}.
For example, for scalar QED, ${\cal D} \phi [g_{\mu \nu}] = {\cal D} (- g)^{1/4} \phi$ and ${\cal D} A_{\mu} [g_{\mu \nu}] = {\cal D} (- g)^{1/4} e_{m}^{\prime ~ \mu} A_{\mu}$, where $e^{m}_{~ \mu}$ is the vierbein.
We change the variables from $\{ \phi'_{i} \}$ in the left hand side to $\{ \phi_{i} \}$ in the right hand side of \cref{eq:matresponse}.
This results in a Jacobian of path integral measure:
\eqs{
\label{eq:Jacob}
{\cal D}  \{ \phi'_{i} \} [g'_{\mu \nu}] \simeq {\cal D} \{ \phi_{i} \} [g_{\mu \nu}] \exp \left(- i \int d^{4} x \sqrt{-g} \omega A_{\rm Jacob} \left( \{ \phi_{i} \}, g_{\mu \nu}; \{ \lambda_{a} \} \right) \right)
}
in the linear variation.
One may evaluate $A_{\rm Jacob}$ by using heat kernel regularization, which is used in Fujikawa's derivation of chiral anomaly~\cite{Fujikawa:1980eg}.
It provides a one-loop contribution to $A_{\rm anom}$, which is proportional to the Weyl tensor squared, the Gauss-Bonnet density, and a gauge field strength squared if $\{ \phi_{i} \}$ is charged.
One can identify $A_{\rm Jacob} = A_{\rm anom}$, which is Fujikawa's derivation of trace anomaly~\cite{Fujikawa:2004cx}.
It follows that the linear variation $A_{\rm lin}$ is given by the {\it quantum} trace of energy-momentum tensor (see \cref{sec:traceT}):
\eqs{
\label{eq:traceT}
A_{\rm lin} =& - \sum_{i} d_{i} ({\rm e.o.m.})_{i} + \left( g_{\mu \nu} T^{\mu \nu} \right)_{\rm class} + A_{\rm anom} \left( \{ \phi_{i} \}, g_{\mu \nu}; \{ \lambda_{a} \} \right) \\
=& g_{\mu \nu} T^{\mu \nu} \left( \{ \phi_{i} \}, g_{\mu \nu}; \{ \lambda_{a} \} \right) \,.
}

In the above discussion, we have taken into account a Jacobian of path integral measure associated with $\{ \phi'_{i} \} \to \{ \phi_{i} \}$ under a background metric.
One also needs to care a Jacobian of path integral measure associated with $g'_{\mu \nu} \to g_{\mu \nu}$ in \cref{eq:gravresponse}.
On the other hand, it is intricate to compute the gravitational Jacobian.
Thus we just assume that it does not give rise to any relevant coupling between scalaron and matter.
For example, in the $R^{2}$ inflation model, the scalaron coupling to matter in \cref{eq:matresponse} reads
\eqs{
S_{\sigma \text{-mat}} = \int d^{4} x \sqrt{-g} \frac{1}{\sqrt{6}} \frac{\sigma}{M_{\rm pl}} g_{\mu \nu} T^{\mu \nu} \,.
}
Our assumption on the gravitational Jacobian reads that it only leads to couplings suppressed by a higher power of $1 / M_{\rm pl}$.
This could be true since the graviton-loop contribution is suppressed by $1 / M_{\rm pl}^{2}$.

In the rest of this paper, we restrict our discussion within the flat spacetime.
The trace of flat-spacetime energy-momentum tensor is enough to evaluate scalaron decay since the scalaron decay amplitude into graviton is further suppressed by $1 / M_{\rm pl}$.

%%%%%%%%%%%%%%%%%%%%%%%%%%%%%%%%%%%%%%%%%%%%%%%%%%%%%%%%%%%%%%%%%%%%%%%%%%%%%%%%%
\section{Trace of energy-momentum tensor \label{sec:flattraceT}}
In the last section we have shown that in the scalaron frame the scalaron couples to matter via the {\it quantum} trace of energy-momentum tensor, by employing the Fujikawa method.
Here we should remark that once we use some regularization, we need to use it throughout, for example, to calculate the renormalization of couplings $\{ \lambda_{a} \}$.
On the other hand, heat kernel regularization in the Fujikawa method is not practical for perturbative renormalization, for which dimensional renormalization is a usual choice.%
\footnote{
Here is a big difference between chiral anomaly and trace anomaly.
Chiral anomaly takes a one-loop exact form~\cite{Adler:1969er, Bardeen:1969md} up to the divergence of some gauge invariant current~\cite{Yonekura:2010mc} due to its topological property, i.e., it counts a number of zero modes in an instanton background~\cite{Atiyah:1968mp, Atiyah:1967ih, Atiyah:1968rj}.
Thus one can use the result from heat kernel regularization even though one uses dimensional regularization for perturbative renormalization.
On the other hand, it does not hold for trace anomaly.
}
In dimensional renormalization, we consider $d = 4 - \epsilon$ dimension instead of four dimension to make loop diagrams finite.
Then we subtract divergences in the four-dimension limit such that counter terms compose solely of poles of $\epsilon$.

In dimensional renormalization, $A_{\rm Jacob}$ does not depend on fields unlike that in the Fujikawa method with heat kernel regularization.
Thus $A_{\rm anom}$ has a different origin in dimensional renormalization.
The trace of energy-momentum tensor takes a form of
\eqs{
\label{eq:flattraceT}
T^{\mu}_{~ \mu} = \lim_{\epsilon \to 0} \left( - \sum_{i} d_{i} ({\rm e.o.m.})_{i} + (T^{\mu}_{~ \mu})_{\rm class} \right) \,.
}
In the right-hand side, a quantity inside the parenthesis is calculated in $d = 4 - \epsilon$ dimension and then taken to the four-dimension limit of $\epsilon \to 0$.
A key observation is that as $\epsilon \to 0$, the second term does not coincide with the four-dimension {\it classical} trace of energy-momentum tensor.
This is because of renormalization (i.e., normal product) of the composite operators such as $F_{\mu \nu}^{2}$ and $|\phi|^{4}$~\cite{Zimmermann:1969jj, Lowenstein:1971jk, Collins:1974da, Breitenlohner:1977hr}.
The renormalization coefficients, including the multiplicative renormalization of bare couplings such as $\lambda$, compose of subtracted poles of $\epsilon$ in the MS or $\overline{\rm MS}$ scheme.
They lead to terms proportional to the $\beta$ function of the renormalized couplings~\cite{Adler:1976zt, Brown:1979pq} such as $\beta_{e} [F_{\mu \nu}^{2}]$ and $\beta_{\lambda} [|\phi|^{4}]$, where the square bracket denotes the renormalized composite operator.
These contributions provide $A_{\rm anom}$.
Also note that $T_{\mu \nu}$ is conserved and thus solely improvement terms arising from non-minimal couplings are renormalized.
Thus $T^{\mu}_{~ \mu}$ is already finite up to renormalization of improvement terms.
In this article we do not go into further detail about renormalization of improvement terms, since it does not change the result at the leading order.

For scalar QED, the Lagrangian density is given by
\eqs{
\label{eq:Lscalar}
{\cal L} = - \frac{1}{4} F_{\mu \nu}^{2} - \frac{1}{2 \xi} (\partial_{\mu} A^{\mu})^{2} + |D_{\mu} \phi|^{2} - m_{s}^{2} |\phi|^{2} - \frac{1}{4} \lambda |\phi|^{4} \,,
}
with $D_{\mu} = \partial_{\mu} - i q e A_{\mu}$ being the gauge covariant derivative for a charge $q$.
We have integrated out the Nakanishi-Lautrup~\cite{Nakanishi:1972sm, Lautrup:1967zz} and (anti-)ghost fields (see \cref{sec:gaugefix}).
$\xi$ is a gauge fixing parameter.%
\footnote{
Note that a gauge fixing parameter $\xi$ is different from a non-minimal coupling $\xi_{\rm grav}$.
}
$d$-dimension flat-spacetime energy-momentum tensor is obtained from \cref{eq:scalarcurvedTmunu} as
\eqs{
T_{\mu \nu} =& - g^{\lambda \kappa} F_{\mu \lambda} F_{\nu \kappa} + 2 D_{\mu} \phi^{*} D_{\nu} \phi - 2 \left( \xi^{c}_{\rm grav} + \frac{\eta}{d - 1} \right) (\partial_{\mu} \partial_{\nu} - g_{\mu \nu} \partial^{2}) |\phi|^{2} \\
& - g_{\mu \nu} \left( - \frac{1}{4} F_{\lambda \kappa}^{2} + |D_{\mu} \phi|^{2} - m_{s}^{2} |\phi|^{2} - \frac{1}{4} \lambda |\phi|^{4} \right) \,.
}
where we rewrite $\xi_{\rm grav} = \xi^{c}_{\rm grav} + \eta / (d - 1)$ with $\xi^{c}_{{\rm grav}} = (d - 2) / (4 (d - 1))$ in $d$ dimension.
We remark that $\eta$ is renormalized in a non-multiplicative manner to make $T_{\mu \nu}$ finite, although we do not go into further detail.
Taking a {\it classical} trace, one finds
\eqs{
\label{eq:scalarclassicalT}
( T^{\mu}_{~ \mu} )_{\rm class} = \epsilon \left( - \frac{1}{4} F_{\mu \nu}^{2} + \frac{1}{4} \lambda |\phi|^{4} \right) + 2 m_{s}^{2} |\phi|^{2} + 2 \eta \partial^{2} |\phi|^{2} + \left(1 - \frac{\epsilon}{2} \right) ({\rm e.o.m}) \,,
}
where the last term with
\eqs{
({\rm e.o.m})  = \phi^{*} \left( D^{2} \phi + m_{s}^{2} \phi + \frac{2}{4} \lambda |\phi|^{2} \phi \right) + \left( D^{2} \phi^{*} + m_{s}^{2} \phi^{*} + \frac{2}{4} \lambda |\phi|^{2} \phi^{*} \right) \phi
}
cancels with $- \sum_{i} d_{i} ({\rm e.o.m.})_{i}$ in \cref{eq:traceT}.
The first term of $( T^{\mu}_{~ \mu} )_{\rm class}$ vanishes at the {\it classical} level as $\epsilon \to 0$, but not at the {\it quantum} level.
This contribution provides $A_{\rm anom}$.

We calculate a $T^{\mu}_{~\mu}$-${\bar A}_{\lambda}$-${\bar A}_{\kappa}$ (${\bar A}_{\mu}$: renormalized gauge field) correlation function in the scalaron frame by using the $\overline{\rm MS}$ scheme.
More specifically, we calculate the amputated amplitude ${\cal M}_{TAA}$ with incoming momentum $k$ through $T^{\mu}_{~\mu}$ and outgoing momentum $k_{1}$ and $k_{2}$ through gauge bosons with helicity $\epsilon_{1}$ and $\epsilon_{2}$, respectively.
For example, in the $R^{2}$ inflation model, the invariant amplitude of scalaron decay into two gauge bosons is given by
\eqs{
{\cal M}_{\rm dec} = \frac{1}{\sqrt{6}} \frac{1}{M_{\rm pl}} {\cal M}_{TAA} \,.
}
\cref{sec:oneloop} is devoted to details of the computations.

For scalar QED (see \cref{sec:scalar}), the leading contribution to ${\cal M}_{TAA}$ arises from the following terms of the trace of energy-momentum tensor:
\eqs{
T^{\mu}_{~ \mu} \supset \frac{1}{6} \frac{q^{2} {\bar e}^{2}}{16 \pi^{2}} {\bar F}_{\mu \nu}^{2} + 2 {\bar m}^{2} |{\bar \phi}|^{2} + 2 {\bar \eta} \partial^{2} |{\bar \phi}|^{2} \,,
}
where the bar denotes the renormalized (not composite) fields and parameters.%
\footnote{
Note that in general ${\bar F}_{\mu \nu}^{2} \neq [F_{\mu \nu}^{2}]$, although they coincide with each other at this order.
}
The first term arises from the gauge kinetic term proportional to $\epsilon$ in \cref{eq:scalarclassicalT}.
Its coefficient is obtained from the leading contribution to the wave function renormalization of the gauge field [see \cref{eq:scalarZ3}].
Meanwhile the leading contribution to the wave function renormalization of the gauge field also determines the leading contribution to the $\beta$ function [see \cref{eq:scalarbeta}] as
\eqs{
\beta_{e} = \frac{1}{3} \frac{q^{2} {\bar e}^{3}}{16 \pi^{2}} \,.
}
The matrix element has two contributions
\eqs{
\label{eq:MTAAscalar}
{\cal M}_{TAA} = {\cal M}_{F^{2}} + {\cal M}_{|\phi|^{2}} \,.
}
The first term arises from the tree-level diagram with the gauge kinetic term inserted:
\eqs{
{\cal M}_{F^{2}} = - \frac{2}{3} \frac{q^{2} {\bar e}^{2}}{16 \pi^{2}} \left( k_{1} \cdot k_{2} \, \epsilon^{*}_{1} \cdot \epsilon^{*}_{2} - k_{2} \cdot \epsilon^{*}_{1} \, k_{1} \cdot \epsilon^{*}_{2} \right) \,.
}
The second term arises from the one-loop diagram with the scalar mass term and $\eta$ term inserted:
\eqs{
{\cal M}_{|\phi|^{2}} = \frac{2}{3} \frac{q^{2} {\bar e}^{2}}{16 \pi^{2}} \frac{{\bar m}^{2} - {\bar \eta} k^{2} }{{\bar m}^{2}} I_{s} \left( \frac{k^{2}}{{\bar m}^{2}} \right) \left( k_{1} \cdot k_{2} \, \epsilon^{*}_{1} \cdot \epsilon^{*}_{2} - k_{2} \cdot \epsilon^{*}_{1} \, k_{1} \cdot \epsilon^{*}_{2} \right) \,,
}
where%
\footnote{
This definition is different from the one in Ref.~\cite{Watanabe:2010vy} by a factor of 6.
}
\eqs{
I_{s} (r) &= 24 \int_{0}^{1} dx \int_{0}^{1 - x} dy \, \frac{x y}{- r x y + 1 - i \epsilon_{\rm ad}} \\
&= 
\begin{cases}
\dfrac{12}{r} \left( - 1 + \dfrac{4}{r} \arcsin^{2} \dfrac{\sqrt{r}}{2} \right) & \text{(for $r < 4$)} \\[6pt]
\dfrac{12}{r} \left( - 1 - \dfrac{4}{r} \left[ {\rm arccosh \,} \dfrac{\sqrt{r}}{2} - i \dfrac{\pi}{2} \right]^{2} \right) & \text{(for $r > 4$)}
\end{cases}
\,.
}
For $r > 4$, one needs to take into account an adiabatic parameter $\epsilon_{\rm ad} > 0$ properly.%
\footnote{
Note that an adiabatic parameter $\epsilon_{\rm ad}$ associated with a Wick rotation is different from $\epsilon = 4 - d$ for dimensional regularization.
}
This arises from the fact that the loop scalar can be real.
Collecting the two contributions, one obtains
\eqs{
{\cal M}_{TAA} = - \frac{2}{3} \frac{q^{2} {\bar e}^{2}}{16 \pi^{2}} \left( 1 - \frac{{\bar m}^{2} - {\bar \eta} k^{2} }{{\bar m}^{2}} I_{s} \left( \frac{k^{2}}{{\bar m}^{2}} \right) \right) \left( k_{1} \cdot k_{2} \, \epsilon^{*}_{1} \cdot \epsilon^{*}_{2} - k_{2} \cdot \epsilon^{*}_{1} \, k_{1} \cdot \epsilon^{*}_{2} \right) \,.
}
We remark that $I_{s} (0) = 1$ and thus a heavy (${\bar m}^2 \gg k^{2}$) scalar does not contribute to ${\cal M}_{TAA}$.
Meanwhile, $I_{s} (\infty) = 0$ and thus a light (${\bar m}^2 \ll k^{2}$) scalar indeed contributes to ${\cal M}_{TAA}$.

It is straightforward to generalize to the case with $N_{s}$ scalars and $N_{f}$ Dirac fermions (see \cref{sec:fermion} for the case with a Dirac fermion) since the quartic and Yukawa coupling do not matter at this order.
The $\beta$ function is given by
\eqs{
\beta_{e} = \frac{1}{3} \left( \sum_{s} q_{s}^{2} + 4 \sum_{f} q_{f}^{2} \right) \frac{{\bar e}^{3}}{16 \pi^{2}} \,.
}
Note that this counts contributions from both {\it heavy} and {\it light} degrees of freedom.

Meanwhile, the matrix element is given by
\eqs{
{\cal M}_{TAA} =& \frac{2}{3} \frac{{\bar e}^{2}}{16 \pi^{2}} \left( \sum_{s} q_{s}^{2} \left( 1 - \frac{{\bar m}_{s}^{2} - {\bar \eta}_{s} k^{2} }{{\bar m}_{s}^{2}}  I_{s} \left( \frac{k^{2}}{{\bar m}_{s}^{2}} \right) \right) + 4  \sum_{f} q_{f}^{2} \left( 1 - I_{f} \left( \frac{k^{2}}{{\bar m}_{f}^{2}} \right) \right) \right) \\ 
& \times \left( k_{1} \cdot k_{2} \, \epsilon^{*}_{1} \cdot \epsilon^{*}_{2} - k_{2} \cdot \epsilon^{*}_{1} \, k_{1} \cdot \epsilon^{*}_{2} \right) \,,
}
where
\eqs{
I_{f} (r) &= 3 \int_{0}^{1} dx \int_{0}^{1 - x} dy \, \frac{- 4 x y + 1}{- r x y + 1 - i \epsilon_{\rm ad}} \\
&= 
\begin{cases}
\dfrac{6}{r} \left( 1 + \left( 1 - \dfrac{4}{r} \right) \arcsin^{2} \dfrac{\sqrt{r}}{2} \right) & \text{(for $r < 4$)} \\[6pt]
\dfrac{6}{r} \left( 1 - \left( 1 - \dfrac{4}{r} \right) \left[ {\rm arccosh \,} \dfrac{\sqrt{r}}{2} - i \dfrac{\pi}{2} \right]^{2} \right) & \text{(for $r > 4$)}
\end{cases}
\,.
}
Here $I_{f} (0) = 1$ and $I_{f} (\infty) = 0$.%
\footnote{
This definition is different from the one in Ref.~\cite{Watanabe:2010vy} by a factor of 3.
}
For $r > 4$, one needs to take into account $\epsilon_{\rm ad}$ properly.
This arises from the fact that the loop fermion can be real.
The matrix element is approximated by
\eqs{
{\cal M}_{TAA} \approx \frac{2}{3} \frac{{\bar e}^{2}}{16 \pi^{2}} \left( \sum_{{\rm light}~s} q_{s}^{2} \left(1 - 12 {\bar \eta}_{s} \right) + 4  \sum_{{\rm light}~f} q_{f}^{2} \right) \left( k_{1} \cdot k_{2} \, \epsilon^{*}_{1} \cdot \epsilon^{*}_{2} - k_{2} \cdot \epsilon^{*}_{1} \, k_{1} \cdot \epsilon^{*}_{2} \right) \,.
}
The summation runs over solely {\it light} scalars or fermions with $m^{2} < k^{2}$.

%%%%%%%%%%%%%%%%%%%%%%%%%%%%%%%%%%%%%%%%%%%%%%%%%%%%%%%%%%%%%%%%%%%%%%%%%%%%%%%%%
\section{Conclusion and remarks \label{sec:concl}}
In this article, we have revisited scalaron decay via the trace of energy-momentum tensor.
In particular we have studied scenarios with a sequestered matter sector, where the trace of energy-momentum tensor gives a dominant contribution to scalaron-matter coupling.
We have shown how trace anomaly arises by employing the Fujikawa method and dimensional renormalization.
For perturbative renormalization beyond the one-loop level, the dimensional renormalization is more convenient than the Fujikawa method.

Trace anomaly plays an important role in ensuring that the trace of energy-momentum tensor is predictive in terms of low-energy effective theory.
We have explicitly calculated the scalaron decay amplitude at the leading order in quantum electrodynamics with scalars and fermions.
The contribution of heavy degrees of freedom through trace anomaly cancels with the one through the mass term, in the heavy mass limit of the scalars and fermions.
It is straightforward to generalize the discussion to quantum chromodynamics.

There are two caveats on the predictability of the trace of energy-momentum tensor: a non-minimal coupling of matter scalars to Ricci curvature; and the renormalization-scale dependence.
They only appear in energy-momentum tensor and thus one cannot be determined its renormalized value through usual experiments unless graviton is involved in a process.
Since a non-minimal coupling is required to renormalize energy-momentum tensor, one should keep it even when one considers a matter sector minimally coupled to gravity.
In addition, it may not be clear how we can see that the scalaron decay amplitude is independent of the renormalization scale, since the trace of energy-momentum tensor is a composite operator.
We will give a detailed discussion on these caveats somewhere else.

\subsection*{Acknowledgement}
The work of A. K. is supported by IBS under the project code, IBS-R018-D1.
A. K. gratefully thanks Heejung Kim, Takumi Kuwahara, and Kazuya Yonekura for valuable discussions.
A. K. thanks Taishi Katsuragawa and Shinya Matsuzaki for discussions on Ref.~\cite{Katsuragawa:2016yir}.
A. K. thanks Yuki Watanabe for discussions on Ref.~\cite{Takeda:2014qma, Watanabe:2010vy}.
A. K. would also like to thank Ryusuke Jinno, Kohei Kamada for encouraging A. K. to work on this paper and providing comments on the manuscript.

\bibliographystyle{./utphys}
\bibliography{ref}

\providecommand{\href}[2]{#2}\begingroup\raggedright\begin{thebibliography}{100}

\bibitem{Brout:1977ix}
R.~Brout, F.~Englert, and E.~Gunzig, ``{The Creation of the Universe as a
  Quantum Phenomenon},''
\href{http://dx.doi.org/10.1016/0003-4916(78)90176-8}{{\em Annals Phys.}
  {\bfseries 115} (1978) 78}.
%%CITATION = APNYA,115,78;%%.

\bibitem{Starobinsky:1980te}
A.~A. Starobinsky, ``{A New Type of Isotropic Cosmological Models Without
  Singularity},'' \href{http://dx.doi.org/10.1016/0370-2693(80)90670-X}{{\em
  Phys. Lett.} {\bfseries B91} (1980) 99--102}.
[,771(1980)].
%%CITATION = PHLTA,B91,99;%%.

\bibitem{Kazanas:1980tx}
D.~Kazanas, ``{Dynamics of the Universe and Spontaneous Symmetry Breaking},''
\href{http://dx.doi.org/10.1086/183361}{{\em Astrophys. J.} {\bfseries 241}
  (1980) L59--L63}.
%%CITATION = ASJOA,241,L59;%%.

\bibitem{Sato:1980yn}
K.~Sato, ``{First Order Phase Transition of a Vacuum and Expansion of the
  Universe},''
{\em Mon. Not. Roy. Astron. Soc.} {\bfseries 195} (1981) 467--479.
%%CITATION = MNRAA,195,467;%%.

\bibitem{Guth:1980zm}
A.~H. Guth, ``{The Inflationary Universe: A Possible Solution to the Horizon
  and Flatness Problems},''
  \href{http://dx.doi.org/10.1103/PhysRevD.23.347}{{\em Phys. Rev.} {\bfseries
  D23} (1981) 347--356}.
[Adv. Ser. Astrophys. Cosmol.3,139(1987)].
%%CITATION = PHRVA,D23,347;%%.

\bibitem{Linde:1981mu}
A.~D. Linde, ``{A New Inflationary Universe Scenario: A Possible Solution of
  the Horizon, Flatness, Homogeneity, Isotropy and Primordial Monopole
  Problems},'' \href{http://dx.doi.org/10.1016/0370-2693(82)91219-9}{{\em Phys.
  Lett.} {\bfseries 108B} (1982) 389--393}.
[Adv. Ser. Astrophys. Cosmol.3,149(1987)].
%%CITATION = PHLTA,108B,389;%%.

\bibitem{Albrecht:1982wi}
A.~Albrecht and P.~J. Steinhardt, ``{Cosmology for Grand Unified Theories with
  Radiatively Induced Symmetry Breaking},''
  \href{http://dx.doi.org/10.1103/PhysRevLett.48.1220}{{\em Phys. Rev. Lett.}
  {\bfseries 48} (1982) 1220--1223}.
[Adv. Ser. Astrophys. Cosmol.3,158(1987)].
%%CITATION = PRLTA,48,1220;%%.

\bibitem{Linde:1983gd}
A.~D. Linde, ``{Chaotic Inflation},''
\href{http://dx.doi.org/10.1016/0370-2693(83)90837-7}{{\em Phys. Lett.}
  {\bfseries 129B} (1983) 177--181}.
%%CITATION = PHLTA,129B,177;%%.

\bibitem{Mukhanov:1981xt}
V.~F. Mukhanov and G.~V. Chibisov, ``{Quantum Fluctuations and a Nonsingular
  Universe},'' {\em JETP Lett.} {\bfseries 33} (1981) 532--535.
[Pisma Zh. Eksp. Teor. Fiz.33,549(1981)].
%%CITATION = JTPLA,33,532;%%.

\bibitem{Mukhanov:1982nu}
V.~F. Mukhanov and G.~V. Chibisov, ``{The Vacuum energy and large scale
  structure of the universe},'' {\em Sov. Phys. JETP} {\bfseries 56} (1982)
  258--265.
[Zh. Eksp. Teor. Fiz.83,475(1982)].
%%CITATION = SPHJA,56,258;%%.

\bibitem{Hawking:1982cz}
S.~W. Hawking, ``{The Development of Irregularities in a Single Bubble
  Inflationary Universe},''
\href{http://dx.doi.org/10.1016/0370-2693(82)90373-2}{{\em Phys. Lett.}
  {\bfseries 115B} (1982) 295}.
%%CITATION = PHLTA,115B,295;%%.

\bibitem{Guth:1982ec}
A.~H. Guth and S.~Y. Pi, ``{Fluctuations in the New Inflationary Universe},''
\href{http://dx.doi.org/10.1103/PhysRevLett.49.1110}{{\em Phys. Rev. Lett.}
  {\bfseries 49} (1982) 1110--1113}.
%%CITATION = PRLTA,49,1110;%%.

\bibitem{Starobinsky:1982ee}
A.~A. Starobinsky, ``{Dynamics of Phase Transition in the New Inflationary
  Universe Scenario and Generation of Perturbations},''
\href{http://dx.doi.org/10.1016/0370-2693(82)90541-X}{{\em Phys. Lett.}
  {\bfseries 117B} (1982) 175--178}.
%%CITATION = PHLTA,117B,175;%%.

\bibitem{Bardeen:1983qw}
J.~M. Bardeen, P.~J. Steinhardt, and M.~S. Turner, ``{Spontaneous Creation of
  Almost Scale - Free Density Perturbations in an Inflationary Universe},''
\href{http://dx.doi.org/10.1103/PhysRevD.28.679}{{\em Phys. Rev.} {\bfseries
  D28} (1983) 679}.
%%CITATION = PHRVA,D28,679;%%.

\bibitem{Mukhanov:1985rz}
V.~F. Mukhanov, ``{Gravitational Instability of the Universe Filled with a
  Scalar Field},'' {\em JETP Lett.} {\bfseries 41} (1985) 493--496.
[Pisma Zh. Eksp. Teor. Fiz.41,402(1985)].
%%CITATION = JTPLA,41,493;%%.

\bibitem{Akrami:2018odb}
{\bfseries Planck} Collaboration, Y.~Akrami {\em et~al.}, ``{Planck 2018
  results. X. Constraints on inflation},''
\href{http://arxiv.org/abs/1807.06211}{{\ttfamily arXiv:1807.06211
  [astro-ph.CO]}}.
%%CITATION = ARXIV:1807.06211;%%.

\bibitem{Martin:2013tda}
J.~Martin, C.~Ringeval, and V.~Vennin, ``{Encyclopædia Inflationaris},''
  \href{http://dx.doi.org/10.1016/j.dark.2014.01.003}{{\em Phys. Dark Univ.}
  {\bfseries 5-6} (2014) 75--235},
\href{http://arxiv.org/abs/1303.3787}{{\ttfamily arXiv:1303.3787
  [astro-ph.CO]}}.
%%CITATION = ARXIV:1303.3787;%%.

\bibitem{Barrow:1983rx}
J.~D. Barrow and A.~C. Ottewill, ``{The Stability of General Relativistic
  Cosmological Theory},''
\href{http://dx.doi.org/10.1088/0305-4470/16/12/022}{{\em J. Phys.} {\bfseries
  A16} (1983) 2757}.
%%CITATION = JPAGA,A16,2757;%%.

\bibitem{Whitt:1984pd}
B.~Whitt, ``{Fourth Order Gravity as General Relativity Plus Matter},''
\href{http://dx.doi.org/10.1016/0370-2693(84)90332-0}{{\em Phys. Lett.}
  {\bfseries 145B} (1984) 176--178}.
%%CITATION = PHLTA,145B,176;%%.

\bibitem{Vilenkin:1985md}
A.~Vilenkin, ``{Classical and Quantum Cosmology of the Starobinsky Inflationary
  Model},''
\href{http://dx.doi.org/10.1103/PhysRevD.32.2511}{{\em Phys. Rev.} {\bfseries
  D32} (1985) 2511}.
%%CITATION = PHRVA,D32,2511;%%.

\bibitem{Mijic:1986iv}
M.~B. Mijic, M.~S. Morris, and W.-M. Suen, ``{The R**2 Cosmology: Inflation
  Without a Phase Transition},''
\href{http://dx.doi.org/10.1103/PhysRevD.34.2934}{{\em Phys. Rev.} {\bfseries
  D34} (1986) 2934}.
%%CITATION = PHRVA,D34,2934;%%.

\bibitem{Barrow:1988xh}
J.~D. Barrow and S.~Cotsakis, ``{Inflation and the Conformal Structure of
  Higher Order Gravity Theories},''
\href{http://dx.doi.org/10.1016/0370-2693(88)90110-4}{{\em Phys. Lett.}
  {\bfseries B214} (1988) 515--518}.
%%CITATION = PHLTA,B214,515;%%.

\bibitem{Matsumura:2016sri}
T.~Matsumura {\em et~al.}, ``{LiteBIRD: Mission Overview and Focal Plane
  Layout},''
\href{http://dx.doi.org/10.1007/s10909-016-1542-8}{{\em J. Low. Temp. Phys.}
  {\bfseries 184} no.~3-4, (2016) 824--831}.
%%CITATION = JLTPA,184,824;%%.

\bibitem{Delabrouille:2017rct}
{\bfseries CORE} Collaboration, J.~Delabrouille {\em et~al.}, ``{Exploring
  cosmic origins with CORE: Survey requirements and mission design},''
  \href{http://dx.doi.org/10.1088/1475-7516/2018/04/014}{{\em JCAP} {\bfseries
  1804} no.~04, (2018) 014},
\href{http://arxiv.org/abs/1706.04516}{{\ttfamily arXiv:1706.04516
  [astro-ph.IM]}}.
%%CITATION = ARXIV:1706.04516;%%.

\bibitem{Abazajian:2016yjj}
{\bfseries CMB-S4} Collaboration, K.~N. Abazajian {\em et~al.}, ``{CMB-S4
  Science Book, First Edition},''
\href{http://arxiv.org/abs/1610.02743}{{\ttfamily arXiv:1610.02743
  [astro-ph.CO]}}.
%%CITATION = ARXIV:1610.02743;%%.

\bibitem{Bezrukov:2011gp}
F.~L. Bezrukov and D.~S. Gorbunov, ``{Distinguishing between R$^2$-inflation
  and Higgs-inflation},''
  \href{http://dx.doi.org/10.1016/j.physletb.2012.06.040}{{\em Phys. Lett.}
  {\bfseries B713} (2012) 365--368},
\href{http://arxiv.org/abs/1111.4397}{{\ttfamily arXiv:1111.4397 [hep-ph]}}.
%%CITATION = ARXIV:1111.4397;%%.

\bibitem{Gorbunov:2010bn}
D.~S. Gorbunov and A.~G. Panin, ``{Scalaron the mighty: producing dark matter
  and baryon asymmetry at reheating},''
  \href{http://dx.doi.org/10.1016/j.physletb.2011.04.067}{{\em Phys. Lett.}
  {\bfseries B700} (2011) 157--162},
\href{http://arxiv.org/abs/1009.2448}{{\ttfamily arXiv:1009.2448 [hep-ph]}}.
%%CITATION = ARXIV:1009.2448;%%.

\bibitem{Gorbunov:2012ij}
D.~S. Gorbunov and A.~G. Panin, ``{Free scalar dark matter candidates in
  $R^2$-inflation: the light, the heavy and the superheavy},''
  \href{http://dx.doi.org/10.1016/j.physletb.2012.10.015}{{\em Phys. Lett.}
  {\bfseries B718} (2012) 15--20},
\href{http://arxiv.org/abs/1201.3539}{{\ttfamily arXiv:1201.3539
  [astro-ph.CO]}}.
%%CITATION = ARXIV:1201.3539;%%.

\bibitem{Buchmuller:2002rq}
W.~Buchmuller, P.~Di~Bari, and M.~Plumacher, ``{Cosmic microwave background,
  matter - antimatter asymmetry and neutrino masses},''
  \href{http://dx.doi.org/10.1016/S0550-3213(02)00737-X,
  10.1016/j.nuclphysb.2007.11.030}{{\em Nucl. Phys.} {\bfseries B643} (2002)
  367--390}, \href{http://arxiv.org/abs/hep-ph/0205349}{{\ttfamily
  arXiv:hep-ph/0205349 [hep-ph]}}.
[Erratum: Nucl. Phys.B793,362(2008)].
%%CITATION = HEP-PH/0205349;%%.

\bibitem{Giudice:2003jh}
G.~F. Giudice, A.~Notari, M.~Raidal, A.~Riotto, and A.~Strumia, ``{Towards a
  complete theory of thermal leptogenesis in the SM and MSSM},''
  \href{http://dx.doi.org/10.1016/j.nuclphysb.2004.02.019}{{\em Nucl. Phys.}
  {\bfseries B685} (2004) 89--149},
\href{http://arxiv.org/abs/hep-ph/0310123}{{\ttfamily arXiv:hep-ph/0310123
  [hep-ph]}}.
%%CITATION = HEP-PH/0310123;%%.

\bibitem{Buchmuller:2005eh}
W.~Buchmuller, R.~D. Peccei, and T.~Yanagida, ``{Leptogenesis as the origin of
  matter},''
  \href{http://dx.doi.org/10.1146/annurev.nucl.55.090704.151558}{{\em Ann. Rev.
  Nucl. Part. Sci.} {\bfseries 55} (2005) 311--355},
\href{http://arxiv.org/abs/hep-ph/0502169}{{\ttfamily arXiv:hep-ph/0502169
  [hep-ph]}}.
%%CITATION = HEP-PH/0502169;%%.

\bibitem{Davidson:2008bu}
S.~Davidson, E.~Nardi, and Y.~Nir, ``{Leptogenesis},''
  \href{http://dx.doi.org/10.1016/j.physrep.2008.06.002}{{\em Phys. Rept.}
  {\bfseries 466} (2008) 105--177},
\href{http://arxiv.org/abs/0802.2962}{{\ttfamily arXiv:0802.2962 [hep-ph]}}.
%%CITATION = ARXIV:0802.2962;%%.

\bibitem{Fukugita:1986hr}
M.~Fukugita and T.~Yanagida, ``{Baryogenesis Without Grand Unification},''
\href{http://dx.doi.org/10.1016/0370-2693(86)91126-3}{{\em Phys. Lett.}
  {\bfseries B174} (1986) 45--47}.
%%CITATION = PHLTA,B174,45;%%.

\bibitem{Gorbunov:2012ns}
D.~Gorbunov and A.~Tokareva, ``{$R^2$-inflation with conformal SM Higgs
  field},'' \href{http://dx.doi.org/10.1088/1475-7516/2013/12/021}{{\em JCAP}
  {\bfseries 1312} (2013) 021},
\href{http://arxiv.org/abs/1212.4466}{{\ttfamily arXiv:1212.4466
  [astro-ph.CO]}}.
%%CITATION = ARXIV:1212.4466;%%.

\bibitem{Kawamura:2006up}
S.~Kawamura {\em et~al.}, ``{The Japanese space gravitational wave antenna
  DECIGO},''
\href{http://dx.doi.org/10.1088/0264-9381/23/8/S17}{{\em Class. Quant. Grav.}
  {\bfseries 23} (2006) S125--S132}.
%%CITATION = CQGRD,23,S125;%%.

\bibitem{Jakubiec:1988ef}
A.~Jakubiec and J.~Kijowski, ``{On Theories of Gravitation With Nonlinear
  Lagrangians},''
\href{http://dx.doi.org/10.1103/PhysRevD.37.1406}{{\em Phys. Rev.} {\bfseries
  D37} (1988) 1406--1409}.
%%CITATION = PHRVA,D37,1406;%%.

\bibitem{Faulkner:2006ub}
T.~Faulkner, M.~Tegmark, E.~F. Bunn, and Y.~Mao, ``{Constraining f(R) Gravity
  as a Scalar Tensor Theory},''
  \href{http://dx.doi.org/10.1103/PhysRevD.76.063505}{{\em Phys. Rev.}
  {\bfseries D76} (2007) 063505},
\href{http://arxiv.org/abs/astro-ph/0612569}{{\ttfamily arXiv:astro-ph/0612569
  [astro-ph]}}.
%%CITATION = ASTRO-PH/0612569;%%.

\bibitem{Accetta:1985du}
F.~S. Accetta, D.~J. Zoller, and M.~S. Turner, ``{Induced Gravity Inflation},''
\href{http://dx.doi.org/10.1103/PhysRevD.31.3046}{{\em Phys. Rev.} {\bfseries
  D31} (1985) 3046}.
%%CITATION = PHRVA,D31,3046;%%.

\bibitem{La:1989za}
D.~La and P.~J. Steinhardt, ``{Extended Inflationary Cosmology},''
  \href{http://dx.doi.org/10.1103/PhysRevLett.62.376,
  10.1103/PhysRevLett.62.1066}{{\em Phys. Rev. Lett.} {\bfseries 62} (1989)
  376}.
[Erratum: Phys. Rev. Lett.62,1066(1989)].
%%CITATION = PRLTA,62,376;%%.

\bibitem{Futamase:1987ua}
T.~Futamase and K.-i. Maeda, ``{Chaotic Inflationary Scenario in Models Having
  Nonminimal Coupling With Curvature},''
\href{http://dx.doi.org/10.1103/PhysRevD.39.399}{{\em Phys. Rev.} {\bfseries
  D39} (1989) 399--404}.
%%CITATION = PHRVA,D39,399;%%.

\bibitem{Salopek:1988qh}
D.~S. Salopek, J.~R. Bond, and J.~M. Bardeen, ``{Designing Density Fluctuation
  Spectra in Inflation},''
\href{http://dx.doi.org/10.1103/PhysRevD.40.1753}{{\em Phys. Rev.} {\bfseries
  D40} (1989) 1753}.
%%CITATION = PHRVA,D40,1753;%%.

\bibitem{Fakir:1990eg}
R.~Fakir and W.~G. Unruh, ``{Improvement on cosmological chaotic inflation
  through nonminimal coupling},''
\href{http://dx.doi.org/10.1103/PhysRevD.41.1783}{{\em Phys. Rev.} {\bfseries
  D41} (1990) 1783--1791}.
%%CITATION = PHRVA,D41,1783;%%.

\bibitem{Takeda:2014qma}
N.~Takeda and Y.~Watanabe, ``{No quasistable scalaron lump forms after $R^2$
  inflation},'' \href{http://dx.doi.org/10.1103/PhysRevD.90.023519}{{\em Phys.
  Rev.} {\bfseries D90} no.~2, (2014) 023519},
\href{http://arxiv.org/abs/1405.3830}{{\ttfamily arXiv:1405.3830
  [astro-ph.CO]}}.
%%CITATION = ARXIV:1405.3830;%%.

\bibitem{Watanabe:2010vy}
Y.~Watanabe, ``{Rate of gravitational inflaton decay via gauge trace
  anomaly},'' \href{http://dx.doi.org/10.1103/PhysRevD.83.043511}{{\em Phys.
  Rev.} {\bfseries D83} (2011) 043511},
\href{http://arxiv.org/abs/1011.3348}{{\ttfamily arXiv:1011.3348 [hep-th]}}.
%%CITATION = ARXIV:1011.3348;%%.

\bibitem{Kannike:2015apa}
K.~Kannike, G.~Hütsi, L.~Pizza, A.~Racioppi, M.~Raidal, A.~Salvio, and
  A.~Strumia, ``{Dynamically Induced Planck Scale and Inflation},''
  \href{http://dx.doi.org/10.1007/JHEP05(2015)065}{{\em JHEP} {\bfseries 05}
  (2015) 065},
\href{http://arxiv.org/abs/1502.01334}{{\ttfamily arXiv:1502.01334
  [astro-ph.CO]}}.
%%CITATION = ARXIV:1502.01334;%%.

\bibitem{Choi:2019osi}
S.-M. Choi, Y.-J. Kang, H.~M. Lee, and K.~Yamashita, ``{Unitary inflaton as
  decaying dark matter},''
\href{http://arxiv.org/abs/1902.03781}{{\ttfamily arXiv:1902.03781 [hep-ph]}}.
%%CITATION = ARXIV:1902.03781;%%.

\bibitem{Ellis:1975ap}
J.~R. Ellis, M.~K. Gaillard, and D.~V. Nanopoulos, ``{A Phenomenological
  Profile of the Higgs Boson},''
\href{http://dx.doi.org/10.1016/0550-3213(76)90382-5}{{\em Nucl. Phys.}
  {\bfseries B106} (1976) 292}.
%%CITATION = NUPHA,B106,292;%%.

\bibitem{Shifman:1979eb}
M.~A. Shifman, A.~I. Vainshtein, M.~B. Voloshin, and V.~I. Zakharov,
  ``{Low-Energy Theorems for Higgs Boson Couplings to Photons},'' {\em Sov. J.
  Nucl. Phys.} {\bfseries 30} (1979) 711--716.
[Yad. Fiz.30,1368(1979)].
%%CITATION = SJNCA,30,711;%%.

\bibitem{Peccei:1977hh}
R.~D. Peccei and H.~R. Quinn, ``{CP Conservation in the Presence of
  Instantons},'' \href{http://dx.doi.org/10.1103/PhysRevLett.38.1440}{{\em
  Phys. Rev. Lett.} {\bfseries 38} (1977) 1440--1443}.
[,328(1977)].
%%CITATION = PRLTA,38,1440;%%.

\bibitem{Peccei:1977ur}
R.~D. Peccei and H.~R. Quinn, ``{Constraints Imposed by CP Conservation in the
  Presence of Instantons},''
\href{http://dx.doi.org/10.1103/PhysRevD.16.1791}{{\em Phys. Rev.} {\bfseries
  D16} (1977) 1791--1797}.
%%CITATION = PHRVA,D16,1791;%%.

\bibitem{Weinberg:1977ma}
S.~Weinberg, ``{A New Light Boson?},''
\href{http://dx.doi.org/10.1103/PhysRevLett.40.223}{{\em Phys. Rev. Lett.}
  {\bfseries 40} (1978) 223--226}.
%%CITATION = PRLTA,40,223;%%.

\bibitem{Wilczek:1977pj}
F.~Wilczek, ``{Problem of Strong p and t Invariance in the Presence of
  Instantons},''
\href{http://dx.doi.org/10.1103/PhysRevLett.40.279}{{\em Phys. Rev. Lett.}
  {\bfseries 40} (1978) 279--282}.
%%CITATION = PRLTA,40,279;%%.

\bibitem{Kim:1979if}
J.~E. Kim, ``{Weak Interaction Singlet and Strong CP Invariance},''
\href{http://dx.doi.org/10.1103/PhysRevLett.43.103}{{\em Phys. Rev. Lett.}
  {\bfseries 43} (1979) 103}.
%%CITATION = PRLTA,43,103;%%.

\bibitem{Shifman:1979if}
M.~A. Shifman, A.~I. Vainshtein, and V.~I. Zakharov, ``{Can Confinement Ensure
  Natural CP Invariance of Strong Interactions?},''
\href{http://dx.doi.org/10.1016/0550-3213(80)90209-6}{{\em Nucl. Phys.}
  {\bfseries B166} (1980) 493--506}.
%%CITATION = NUPHA,B166,493;%%.

\bibitem{Zhitnitsky:1980tq}
A.~R. Zhitnitsky, ``{On Possible Suppression of the Axion Hadron Interactions.
  (In Russian)},'' {\em Sov. J. Nucl. Phys.} {\bfseries 31} (1980) 260.
[Yad. Fiz.31,497(1980)].
%%CITATION = SJNCA,31,260;%%.

\bibitem{Dine:1981rt}
M.~Dine, W.~Fischler, and M.~Srednicki, ``{A Simple Solution to the Strong CP
  Problem with a Harmless Axion},''
\href{http://dx.doi.org/10.1016/0370-2693(81)90590-6}{{\em Phys. Lett.}
  {\bfseries B104} (1981) 199--202}.
%%CITATION = PHLTA,B104,199;%%.

\bibitem{Katsuragawa:2016yir}
T.~Katsuragawa and S.~Matsuzaki, ``{Dark matter in modified gravity?},''
  \href{http://dx.doi.org/10.1103/PhysRevD.95.044040}{{\em Phys. Rev.}
  {\bfseries D95} no.~4, (2017) 044040},
\href{http://arxiv.org/abs/1610.01016}{{\ttfamily arXiv:1610.01016 [gr-qc]}}.
%%CITATION = ARXIV:1610.01016;%%.

\bibitem{Callan:1970ze}
C.~G. Callan, Jr., S.~R. Coleman, and R.~Jackiw, ``{A New improved energy -
  momentum tensor},''
\href{http://dx.doi.org/10.1016/0003-4916(70)90394-5}{{\em Annals Phys.}
  {\bfseries 59} (1970) 42--73}.
%%CITATION = APNYA,59,42;%%.

\bibitem{Coleman:1970je}
S.~R. Coleman and R.~Jackiw, ``{Why dilatation generators do not generate
  dilatations?},''
\href{http://dx.doi.org/10.1016/0003-4916(71)90153-9}{{\em Annals Phys.}
  {\bfseries 67} (1971) 552--598}.
%%CITATION = APNYA,67,552;%%.

\bibitem{Freedman:1974gs}
D.~Z. Freedman, I.~J. Muzinich, and E.~J. Weinberg, ``{On the Energy-Momentum
  Tensor in Gauge Field Theories},''
\href{http://dx.doi.org/10.1016/0003-4916(74)90448-5}{{\em Annals Phys.}
  {\bfseries 87} (1974) 95}.
%%CITATION = APNYA,87,95;%%.

\bibitem{Freedman:1974ze}
D.~Z. Freedman and E.~J. Weinberg, ``{The Energy-Momentum Tensor in Scalar and
  Gauge Field Theories},''
\href{http://dx.doi.org/10.1016/0003-4916(74)90040-2}{{\em Annals Phys.}
  {\bfseries 87} (1974) 354}.
%%CITATION = APNYA,87,354;%%.

\bibitem{Collins:1976vm}
J.~C. Collins, ``{The Energy-Momentum Tensor Revisited},''
\href{http://dx.doi.org/10.1103/PhysRevD.14.1965}{{\em Phys. Rev.} {\bfseries
  D14} (1976) 1965}.
%%CITATION = PHRVA,D14,1965;%%.

\bibitem{Nielsen:1977sy}
N.~K. Nielsen, ``{The Energy Momentum Tensor in a Nonabelian Quark Gluon
  Theory},''
\href{http://dx.doi.org/10.1016/0550-3213(77)90040-2}{{\em Nucl. Phys.}
  {\bfseries B120} (1977) 212--220}.
%%CITATION = NUPHA,B120,212;%%.

\bibitem{Adler:1976zt}
S.~L. Adler, J.~C. Collins, and A.~Duncan, ``{Energy-Momentum-Tensor Trace
  Anomaly in Spin 1/2 Quantum Electrodynamics},''
  \href{http://dx.doi.org/10.1103/PhysRevD.15.1712}{{\em Phys. Rev.} {\bfseries
  D15} (1977) 1712}.
[,318(1976)].
%%CITATION = PHRVA,D15,1712;%%.

\bibitem{Collins:1976yq}
J.~C. Collins, A.~Duncan, and S.~D. Joglekar, ``{Trace and Dilatation Anomalies
  in Gauge Theories},''
\href{http://dx.doi.org/10.1103/PhysRevD.16.438}{{\em Phys. Rev.} {\bfseries
  D16} (1977) 438--449}.
%%CITATION = PHRVA,D16,438;%%.

\bibitem{Brown:1979pq}
L.~S. Brown, ``{Dimensional Regularization of Composite Operators in Scalar
  Field Theory},''
\href{http://dx.doi.org/10.1016/0003-4916(80)90377-2}{{\em Annals Phys.}
  {\bfseries 126} (1980) 135}.
%%CITATION = APNYA,126,135;%%.

\bibitem{Brown:1980qq}
L.~S. Brown and J.~C. Collins, ``{Dimensional Renormalization of Scalar Field
  Theory in Curved Space-time},''
\href{http://dx.doi.org/10.1016/0003-4916(80)90232-8}{{\em Annals Phys.}
  {\bfseries 130} (1980) 215}.
%%CITATION = APNYA,130,215;%%.

\bibitem{Hathrell:1981zb}
S.~J. Hathrell, ``{Trace Anomalies and $\lambda \phi^4$ Theory in Curved
  Space},''
\href{http://dx.doi.org/10.1016/0003-4916(82)90008-2}{{\em Annals Phys.}
  {\bfseries 139} (1982) 136}.
%%CITATION = APNYA,139,136;%%.

\bibitem{Hathrell:1981gz}
S.~J. Hathrell, ``{Trace Anomalies and {QED} in Curved Space},''
\href{http://dx.doi.org/10.1016/0003-4916(82)90227-5}{{\em Annals Phys.}
  {\bfseries 142} (1982) 34}.
%%CITATION = APNYA,142,34;%%.

\bibitem{Capper:1974ed}
D.~M. Capper, M.~J. Duff, and L.~Halpern, ``{Photon corrections to the graviton
  propagator},''
\href{http://dx.doi.org/10.1103/PhysRevD.10.461}{{\em Phys. Rev.} {\bfseries
  D10} (1974) 461--467}.
%%CITATION = PHRVA,D10,461;%%.

\bibitem{Deser:1974cz}
S.~Deser and P.~van Nieuwenhuizen, ``{One Loop Divergences of Quantized
  Einstein-Maxwell Fields},''
\href{http://dx.doi.org/10.1103/PhysRevD.10.401}{{\em Phys. Rev.} {\bfseries
  D10} (1974) 401}.
%%CITATION = PHRVA,D10,401;%%.

\bibitem{Capper:1974ic}
D.~M. Capper and M.~J. Duff, ``{Trace anomalies in dimensional
  regularization},''
\href{http://dx.doi.org/10.1007/BF02748300}{{\em Nuovo Cim.} {\bfseries A23}
  (1974) 173--183}.
%%CITATION = NUCIA,A23,173;%%.

\bibitem{Capper:1973mv}
D.~M. Capper and M.~J. Duff, ``{THE ONE LOOP NEUTRINO CONTRIBUTION TO THE
  GRAVITON PROPAGATOR},''
\href{http://dx.doi.org/10.1016/0550-3213(74)90582-3}{{\em Nucl. Phys.}
  {\bfseries B82} (1974) 147--154}.
%%CITATION = NUPHA,B82,147;%%.

\bibitem{Dowker:1976zf}
J.~S. Dowker and R.~Critchley, ``{The Stress Tensor Conformal Anomaly for
  Scalar and Spinor Fields},''
\href{http://dx.doi.org/10.1103/PhysRevD.16.3390}{{\em Phys. Rev.} {\bfseries
  D16} (1977) 3390}.
%%CITATION = PHRVA,D16,3390;%%.

\bibitem{Brown:1976wc}
L.~S. Brown, ``{Stress Tensor Trace Anomaly in a Gravitational Metric: Scalar
  Fields},''
\href{http://dx.doi.org/10.1103/PhysRevD.15.1469}{{\em Phys. Rev.} {\bfseries
  D15} (1977) 1469}.
%%CITATION = PHRVA,D15,1469;%%.

\bibitem{Christensen:1977jc}
S.~M. Christensen and S.~A. Fulling, ``{Trace Anomalies and the Hawking
  Effect},''
\href{http://dx.doi.org/10.1103/PhysRevD.15.2088}{{\em Phys. Rev.} {\bfseries
  D15} (1977) 2088--2104}.
%%CITATION = PHRVA,D15,2088;%%.

\bibitem{Brown:1977pq}
L.~S. Brown and J.~P. Cassidy, ``{Stress Tensor Trace Anomaly in a
  Gravitational Metric: General Theory, Maxwell Field},''
\href{http://dx.doi.org/10.1103/PhysRevD.15.2810}{{\em Phys. Rev.} {\bfseries
  D15} (1977) 2810}.
%%CITATION = PHRVA,D15,2810;%%.

\bibitem{Duff:1977ay}
M.~J. Duff, ``{Observations on Conformal Anomalies},''
\href{http://dx.doi.org/10.1016/0550-3213(77)90410-2}{{\em Nucl. Phys.}
  {\bfseries B125} (1977) 334--348}.
%%CITATION = NUPHA,B125,334;%%.

\bibitem{Bunch:1978yq}
T.~S. Bunch and P.~C.~W. Davies, ``{Quantum Field Theory in de Sitter Space:
  Renormalization by Point Splitting},''
\href{http://dx.doi.org/10.1098/rspa.1978.0060}{{\em Proc. Roy. Soc. Lond.}
  {\bfseries A360} (1978) 117--134}.
%%CITATION = PRSLA,A360,117;%%.

\bibitem{Duff:1993wm}
M.~J. Duff, ``{Twenty years of the Weyl anomaly},''
  \href{http://dx.doi.org/10.1088/0264-9381/11/6/004}{{\em Class. Quant. Grav.}
  {\bfseries 11} (1994) 1387--1404},
\href{http://arxiv.org/abs/hep-th/9308075}{{\ttfamily arXiv:hep-th/9308075
  [hep-th]}}.
%%CITATION = HEP-TH/9308075;%%.

\bibitem{Randall:1998uk}
L.~Randall and R.~Sundrum, ``{Out of this world supersymmetry breaking},''
  \href{http://dx.doi.org/10.1016/S0550-3213(99)00359-4}{{\em Nucl. Phys.}
  {\bfseries B557} (1999) 79--118},
\href{http://arxiv.org/abs/hep-th/9810155}{{\ttfamily arXiv:hep-th/9810155
  [hep-th]}}.
%%CITATION = HEP-TH/9810155;%%.

\bibitem{Giudice:1998xp}
G.~F. Giudice, M.~A. Luty, H.~Murayama, and R.~Rattazzi, ``{Gaugino mass
  without singlets},''
  \href{http://dx.doi.org/10.1088/1126-6708/1998/12/027}{{\em JHEP} {\bfseries
  12} (1998) 027},
\href{http://arxiv.org/abs/hep-ph/9810442}{{\ttfamily arXiv:hep-ph/9810442
  [hep-ph]}}.
%%CITATION = HEP-PH/9810442;%%.

\bibitem{Endo:2007ih}
M.~Endo, F.~Takahashi, and T.~T. Yanagida, ``{Anomaly-induced inflaton decay
  and gravitino-overproduction problem},''
  \href{http://dx.doi.org/10.1016/j.physletb.2007.09.019}{{\em Phys. Lett.}
  {\bfseries B658} (2008) 236--240},
\href{http://arxiv.org/abs/hep-ph/0701042}{{\ttfamily arXiv:hep-ph/0701042
  [hep-ph]}}.
%%CITATION = HEP-PH/0701042;%%.

\bibitem{Endo:2007sz}
M.~Endo, F.~Takahashi, and T.~T. Yanagida, ``{Inflaton Decay in
  Supergravity},'' \href{http://dx.doi.org/10.1103/PhysRevD.76.083509}{{\em
  Phys. Rev.} {\bfseries D76} (2007) 083509},
\href{http://arxiv.org/abs/0706.0986}{{\ttfamily arXiv:0706.0986 [hep-ph]}}.
%%CITATION = ARXIV:0706.0986;%%.

\bibitem{Terada:2014uia}
T.~Terada, Y.~Watanabe, Y.~Yamada, and J.~Yokoyama, ``{Reheating processes
  after Starobinsky inflation in old-minimal supergravity},''
  \href{http://dx.doi.org/10.1007/JHEP02(2015)105}{{\em JHEP} {\bfseries 02}
  (2015) 105},
\href{http://arxiv.org/abs/1411.6746}{{\ttfamily arXiv:1411.6746 [hep-ph]}}.
%%CITATION = ARXIV:1411.6746;%%.

\bibitem{Falls:2018olk}
K.~Falls and M.~Herrero-Valea, ``{Frame (In)equivalence in Quantum Field Theory
  and Cosmology},''
\href{http://arxiv.org/abs/1812.08187}{{\ttfamily arXiv:1812.08187 [hep-th]}}.
%%CITATION = ARXIV:1812.08187;%%.

\bibitem{Fujikawa:1980vr}
K.~Fujikawa, ``{Comment on Chiral and Conformal Anomalies},''
\href{http://dx.doi.org/10.1103/PhysRevLett.44.1733}{{\em Phys. Rev. Lett.}
  {\bfseries 44} (1980) 1733}.
%%CITATION = PRLTA,44,1733;%%.

\bibitem{Fujikawa:1980rc}
K.~Fujikawa, ``{Energy Momentum Tensor in Quantum Field Theory},''
\href{http://dx.doi.org/10.1103/PhysRevD.23.2262}{{\em Phys. Rev.} {\bfseries
  D23} (1981) 2262}.
%%CITATION = PHRVA,D23,2262;%%.

\bibitem{Fujikawa:1993xv}
K.~Fujikawa, ``{A nondiagramatic calculation of one loop beta function in
  QCD},''
\href{http://dx.doi.org/10.1103/PhysRevD.48.3922}{{\em Phys. Rev.} {\bfseries
  D48} (1993) 3922--3924}.
%%CITATION = PHRVA,D48,3922;%%.

\bibitem{Fujikawa:2004cx}
K.~Fujikawa and H.~Suzuki,
  \href{http://dx.doi.org/10.1093/acprof:oso/9780198529132.001.0001}{{\em {Path
  integrals and quantum anomalies}}}.
\newblock
2004.
\newblock
%%CITATION = INSPIRE-655753;%%.

\bibitem{Bollini:1972ui}
C.~G. Bollini and J.~J. Giambiagi, ``{Dimensional Renormalization: The Number
  of Dimensions as a Regularizing Parameter},''
\href{http://dx.doi.org/10.1007/BF02895558}{{\em Nuovo Cim.} {\bfseries B12}
  (1972) 20--26}.
%%CITATION = NUCIA,B12,20;%%.

\bibitem{Ashmore:1972uj}
J.~F. Ashmore, ``{A Method of Gauge Invariant Regularization},''
\href{http://dx.doi.org/10.1007/BF02824407}{{\em Lett. Nuovo Cim.} {\bfseries
  4} (1972) 289--290}.
%%CITATION = NCLTA,4,289;%%.

\bibitem{tHooft:1972tcz}
G.~'t~Hooft and M.~J.~G. Veltman, ``{Regularization and Renormalization of
  Gauge Fields},''
\href{http://dx.doi.org/10.1016/0550-3213(72)90279-9}{{\em Nucl. Phys.}
  {\bfseries B44} (1972) 189--213}.
%%CITATION = NUPHA,B44,189;%%.

\bibitem{Kolb:1990vq}
E.~W. Kolb and M.~S. Turner, ``{The Early Universe},''
{\em Front. Phys.} {\bfseries 69} (1990) 1--547.
%%CITATION = FRPHA,69,1;%%.

\bibitem{Fujikawa:1980eg}
K.~Fujikawa, ``{Path Integral for Gauge Theories with Fermions},''
  \href{http://dx.doi.org/10.1103/PhysRevD.21.2848,
  10.1103/PhysRevD.22.1499}{{\em Phys. Rev.} {\bfseries D21} (1980) 2848}.
[Erratum: Phys. Rev.D22,1499(1980)].
%%CITATION = PHRVA,D21,2848;%%.

\bibitem{Adler:1969er}
S.~L. Adler and W.~A. Bardeen, ``{Absence of higher order corrections in the
  anomalous axial vector divergence equation},''
  \href{http://dx.doi.org/10.1103/PhysRev.182.1517}{{\em Phys. Rev.} {\bfseries
  182} (1969) 1517--1536}.
[,268(1969)].
%%CITATION = PHRVA,182,1517;%%.

\bibitem{Bardeen:1969md}
W.~A. Bardeen, ``{Anomalous Ward identities in spinor field theories},''
\href{http://dx.doi.org/10.1103/PhysRev.184.1848}{{\em Phys. Rev.} {\bfseries
  184} (1969) 1848--1857}.
%%CITATION = PHRVA,184,1848;%%.

\bibitem{Yonekura:2010mc}
K.~Yonekura, ``{Notes on Operator Equations of Supercurrent Multiplets and
  Anomaly Puzzle in Supersymmetric Field Theories},''
  \href{http://dx.doi.org/10.1007/JHEP09(2010)049}{{\em JHEP} {\bfseries 09}
  (2010) 049},
\href{http://arxiv.org/abs/1004.1296}{{\ttfamily arXiv:1004.1296 [hep-th]}}.
%%CITATION = ARXIV:1004.1296;%%.

\bibitem{Atiyah:1968mp}
M.~F. Atiyah and I.~M. Singer, ``{The Index of elliptic operators. 1},''
\href{http://dx.doi.org/10.2307/1970715}{{\em Annals Math.} {\bfseries 87}
  (1968) 484--530}.
%%CITATION = ANMAA,87,484;%%.

\bibitem{Atiyah:1967ih}
M.~F. Atiyah and I.~M. Singer, ``{The Index of elliptic operators. 3.},''
\href{http://dx.doi.org/10.2307/1970717}{{\em Annals Math.} {\bfseries 87}
  (1968) 546--604}.
%%CITATION = ANMAA,87,546;%%.

\bibitem{Atiyah:1968rj}
M.~F. Atiyah and G.~B. Segal, ``{The Index of elliptic operators. 2.},''
\href{http://dx.doi.org/10.2307/1970716}{{\em Annals Math.} {\bfseries 87}
  (1968) 531--545}.
%%CITATION = ANMAA,87,531;%%.

\bibitem{Zimmermann:1969jj}
W.~Zimmermann, ``{Convergence of Bogolyubov's method of renormalization in
  momentum space},'' \href{http://dx.doi.org/10.1007/BF01645676}{{\em Commun.
  Math. Phys.} {\bfseries 15} (1969) 208--234}.
[Lect. Notes Phys.558,217(2000)].
%%CITATION = CMPHA,15,208;%%.

\bibitem{Lowenstein:1971jk}
J.~H. Lowenstein, ``{Differential vertex operations in Lagrangian field
  theory},''
\href{http://dx.doi.org/10.1007/BF01907030}{{\em Commun. Math. Phys.}
  {\bfseries 24} (1971) 1--21}.
%%CITATION = CMPHA,24,1;%%.

\bibitem{Collins:1974da}
J.~C. Collins, ``{Normal Products in Dimensional Regularization},''
\href{http://dx.doi.org/10.1016/S0550-3213(75)80010-1}{{\em Nucl. Phys.}
  {\bfseries B92} (1975) 477--506}.
%%CITATION = NUPHA,B92,477;%%.

\bibitem{Breitenlohner:1977hr}
P.~Breitenlohner and D.~Maison, ``{Dimensional Renormalization and the Action
  Principle},''
\href{http://dx.doi.org/10.1007/BF01609069}{{\em Commun. Math. Phys.}
  {\bfseries 52} (1977) 11--38}.
%%CITATION = CMPHA,52,11;%%.

\bibitem{Nakanishi:1972sm}
N.~Nakanishi, ``{Massive vector field and electromagnetic field in the landau
  gauge},''
\href{http://dx.doi.org/10.1103/PhysRevD.5.1324}{{\em Phys. Rev.} {\bfseries
  D5} (1972) 1324--1330}.
%%CITATION = PHRVA,D5,1324;%%.

\bibitem{Lautrup:1967zz}
B.~Lautrup, ``{CANONICAL QUANTUM ELECTRODYNAMICS IN COVARIANT GAUGES},''
{\em Kong. Dan. Vid. Sel. Mat. Fys. Med.} {\bfseries 35} no.~11, (1967) .
%%CITATION = KDVSA,35,;%%.

\bibitem{Becchi:1974xu}
C.~Becchi, A.~Rouet, and R.~Stora, ``{The Abelian Higgs-Kibble Model. Unitarity
  of the S Operator},''
\href{http://dx.doi.org/10.1016/0370-2693(74)90058-6}{{\em Phys. Lett.}
  {\bfseries 52B} (1974) 344--346}.
%%CITATION = PHLTA,52B,344;%%.

\bibitem{Becchi:1974md}
C.~Becchi, A.~Rouet, and R.~Stora, ``{Renormalization of the Abelian
  Higgs-Kibble Model},''
\href{http://dx.doi.org/10.1007/BF01614158}{{\em Commun. Math. Phys.}
  {\bfseries 42} (1975) 127--162}.
%%CITATION = CMPHA,42,127;%%.

\bibitem{Becchi:1975nq}
C.~Becchi, A.~Rouet, and R.~Stora, ``{Renormalization of Gauge Theories},''
\href{http://dx.doi.org/10.1016/0003-4916(76)90156-1}{{\em Annals Phys.}
  {\bfseries 98} (1976) 287--321}.
%%CITATION = APNYA,98,287;%%.

\bibitem{Tyutin:1975qk}
I.~V. Tyutin, ``{Gauge Invariance in Field Theory and Statistical Physics in
  Operator Formalism},''
\href{http://arxiv.org/abs/0812.0580}{{\ttfamily arXiv:0812.0580 [hep-th]}}.
%%CITATION = ARXIV:0812.0580;%%.

\bibitem{Kugo:1977zq}
T.~Kugo and I.~Ojima, ``{Manifestly Covariant Canonical Formulation of
  Yang-Mills Field Theories: Physical State Subsidiary Conditions and Physical
  S Matrix Unitarity},''
\href{http://dx.doi.org/10.1016/0370-2693(78)90765-7}{{\em Phys. Lett.}
  {\bfseries 73B} (1978) 459--462}.
%%CITATION = PHLTA,73B,459;%%.

\bibitem{Kugo:1977yx}
T.~Kugo and I.~Ojima, ``{Manifestly Covariant Canonical Formulation of
  Yang-Mills Field Theories. 1. The Case of Yang-Mills Fields of Higgs-Kibble
  Type in Landau Gauge},''
\href{http://dx.doi.org/10.1143/PTP.60.1869}{{\em Prog. Theor. Phys.}
  {\bfseries 60} (1978) 1869}.
%%CITATION = PTPKA,60,1869;%%.

\bibitem{Kugo:1979gm}
T.~Kugo and I.~Ojima, ``{Local Covariant Operator Formalism of Nonabelian Gauge
  Theories and Quark Confinement Problem},''
\href{http://dx.doi.org/10.1143/PTPS.66.1}{{\em Prog. Theor. Phys. Suppl.}
  {\bfseries 66} (1979) 1--130}.
%%CITATION = PTPSA,66,1;%%.

\bibitem{Passarino:1978jh}
G.~Passarino and M.~J.~G. Veltman, ``{One Loop Corrections for e+ e-
  Annihilation Into mu+ mu- in the Weinberg Model},''
\href{http://dx.doi.org/10.1016/0550-3213(79)90234-7}{{\em Nucl. Phys.}
  {\bfseries B160} (1979) 151--207}.
%%CITATION = NUPHA,B160,151;%%.

\bibitem{tHooft:1978jhc}
G.~'t~Hooft and M.~J.~G. Veltman, ``{Scalar One Loop Integrals},''
\href{http://dx.doi.org/10.1016/0550-3213(79)90605-9}{{\em Nucl. Phys.}
  {\bfseries B153} (1979) 365--401}.
%%CITATION = NUPHA,B153,365;%%.

\bibitem{Logan:1999if}
H.~E. Logan, {\em {Radiative corrections to the Z b anti-b vertex and
  constraints on extended Higgs sectors}}.
\newblock PhD thesis, UC, Santa Cruz, 1999.
\newblock \href{http://arxiv.org/abs/hep-ph/9906332}{{\ttfamily
  arXiv:hep-ph/9906332 [hep-ph]}}.
\newblock
\url{http://wwwlib.umi.com/dissertations/fullcit?p9940272}.
\newblock
%%CITATION = HEP-PH/9906332;%%.

\end{thebibliography}\endgroup

\newpage
\appendix

%%%%%%%%%%%%%%%%%%%%%%%%%%%%%%%%%%%%%%%%%%%%%%%%%%%%%%%%%%%%%%%%%%%%%%%%%%%%%%%%%
\section{Gauge fixing term \label{sec:gaugefix}}
In this section we discuss the gauge fixing term $S_{\rm fix}$ in non-Abelian gauge theory, while we consider Abelian gauge theory (QED) in the main text.
The gauge fixing term takes a Becchi-Rouet-Stora-Tyutin (BRST) form~\cite{Becchi:1974xu, Becchi:1974md, Becchi:1975nq, Tyutin:1975qk} of
\eqs{
S_{\rm fix} = \int d^{d} x \sqrt{- g} \left( \frac{\xi}{2} B^{a} B^{a} - g^{\mu \nu} \nabla_{\mu} B^{a} A^{a}_{~ \nu} + g^{\mu \nu} \nabla_{\mu} {\bar c}^{a} D_{\nu} c^{a} \right) \,,
}
with $\xi$ being a gauge fixing parameter.
The superscript $a$ runs over gauge group generators $T^{a}$ [$T^{a} = {\mathbb I}$ (identity matrix) in QED].
$D_{\mu}$ is the gauge and diffeomorphism covariant derivative, while $\nabla_{\mu}$ is the diffeomorphism (not gauge) covariant derivative.
We have introduced a bosonic auxiliary Nakanishi-Lautrup field $B^{a} = B^{a \dagger}$, fermionic (ghost and anti-ghost) fields, $c^{a} = c^{a \dagger}$ and $\bar c^{a} = - {\bar c}^{a \dagger}$.

The BRST transformation is defined by the following fermionic global transformation:%
\eqs{
& Q A_{\mu} = D_{\mu} c \,, \\
& Q c = \frac{i}{2} e [c, c] \,, \\
& Q {\bar c} = B \,, \\
& Q B = 0 \,,
}
with $e$ being a gauge coupling.
We have used the matrix notation of $A_{\mu} = A^{a}_{~ \mu} T^{a}$ and $D_{\mu} c = \partial_{\mu} c - i e [c, A]$, and so on.
These are understood as $[Q, A_{\mu}] = i D_{\mu} c$ (commutator), $\{Q, {\bar c}\} = i B$ (anti-commutator), and so on in the operator formalism with $Q^{\dagger} = Q$.
An operator or state is called BRST closed when it vanishes under the BRST transformation.
Gauge invariant operators, such as a gauge invariant part of an action and its contribution to energy-momentum tensor [see eq.~\cref{eq:scalarcurvedTmunu}], are BRST closed.
Meanwhile an operator or state is called BRST exact when it can be written as the BRST transformation of some operator or state.
Notably the gauge fixing term is BRST exact:
\begin{eqnarray}
S_{\rm fix} = \int d^{d} x \sqrt{- g} Q \left( \frac{\xi}{2} {\bar c}^{a} B^{a} - g^{\mu \nu} \nabla_{\mu} {\bar c}^{a} A^{a}_{~ \nu} \right) \,.
\end{eqnarray}
$S_{\rm fix}$ contribution to the energy-momentum tensor is also BRST exact:
\eqs{
T^{\rm fix}_{\mu \nu} = Q \left( - \nabla_{\mu} {\bar c}^{a} A^{a}_{~ \nu} - \nabla_{\nu} {\bar c}^{a} A^{a}_{~ \mu} - g_{\mu \nu} \left( \frac{\xi}{2} {\bar c}^{a} B^{a} - g^{\lambda \kappa} \nabla_{\lambda} {\bar c}^{a} A^{a}_{~ \kappa} \right) \right) \,.
}

One can see that the BRST transformation is nilpotent: $Q^{2} = 0$.
Thus a BRST-exact operator or state is BRST closed.
We can introduce an equivalence class on the set of BRST-closed operators or states ${\cal H}_{\rm closed}$ as ${\cal H}_{\rm closed} \sim {\cal H}_{\rm closed} + {\cal H}_{\rm exact}$ with the set of BRST-exact operators or states ${\cal H}_{\rm exact} \subset {\cal H}_{\rm closed}$.
The physical operator or state is defined by the quotient set of ${\cal H}_{\rm closed} / {\cal H}_{\rm exact}$~\cite{Kugo:1977zq, Kugo:1977yx, Kugo:1979gm}.
Since $T^{\rm fix}_{\mu \nu}$ is BRST exact, one can chose a physical representative such that $T^{\rm fix}_{\mu \nu} = 0$.

%%%%%%%%%%%%%%%%%%%%%%%%%%%%%%%%%%%%%%%%%%%%%%%%%%%%%%%%%%%%%%%%%%%%%%%%%%%%%%%%%
\section{Path integral derivation of \cref{eq:traceT} \label{sec:traceT}}
We consider a correlation function in the path integral formalism:
\eqs{
& \int {\cal D} {\{ \phi_{i} \}} [g'_{\mu \nu}] \exp \left( i S_{\rm mat} \left[ \{ \phi_{i} \}, g'_{\mu \nu} ; \{ \lambda_{a} \} \right] \right) \prod {\{ \phi_{i} \}} \\
& \simeq \left( 1 + \int d^{4} x \sqrt{- g} \omega g_{\mu \nu} \frac{2}{\sqrt{- g}} \frac{\delta}{\delta g_{\mu \nu}} \right) \int {\cal D} {\{ \phi_{i} \}} [g_{\mu \nu}] \exp \left( i S_{\rm mat} \left[ \{ \phi_{i} \}, g_{\mu \nu} ; \{ \lambda_{a} \} \right] \right) \prod {\{ \phi_{i} \}} \,.
}
Meanwhile,
\eqs{
& \int {\cal D} {\{ \phi_{i} \}} [g'_{\mu \nu}] \exp \left( i S_{\rm mat} \left[ \{ \phi_{i} \}, g'_{\mu \nu} ; \{ \lambda_{a} \} \right] \right) \prod {\{ \phi_{i} \}} \\
& = \int {\cal D} {\{ \phi'_{i} \}} [g'_{\mu \nu}] \exp \left( i S_{\rm mat} \left[ \{ \phi'_{i} \}, g'_{\mu \nu} ; \{ \lambda_{a} \} \right] \right) \prod {\{ \phi'_{i} \}} \\
& \simeq \int {\cal D} {\{ \phi_{i} \}} \exp \left( i S_{\rm mat} \left[ \{ \phi_{i} \}, g_{\mu \nu} ; \{ \lambda_{a} \} \right] \right) \left( 1 + i \int d^{4} x \sqrt{-g} \omega \left[ \sum_{i} d_{i} ({\rm e.o.m.})_{i} \right. \right. \\
& \left. \left. - g_{\mu \nu} T^{\mu \nu} \left( \{ \phi_{i} \}, g_{\mu \nu}; \{ \lambda_{a} \} \right) -  A_{\rm Jacob} \left( \{ \phi_{i} \}, g_{\mu \nu} ; \{ \lambda_{a} \} \right)  \right] - \int d^{4} x \omega \sum_{i} \frac{d_{i} \phi_{i}}{\sqrt{-g}} \frac{\delta}{\delta \phi_{i}} \right) \prod {\{ \phi_{i} \}} \,.
}
In the first equality, we change a notation of the integration variable $\{ \phi_{i} \}$, which has no physical effect.
From this Ward-Takahashi identity, one finds
\eqs{
- g_{\mu \nu} T^{\mu \nu} = \sum_{i} d_{i} ({\rm e.o.m.})_{i} - (g_{\mu \nu} T^{\mu \nu} )_{\rm class} - A_{\rm Jacob} \left( \{ \phi_{i} \}, g_{\mu \nu} ; \{ \lambda_{a} \} \right) \,,
}
by ignoring the contact terms.

%%%%%%%%%%%%%%%%%%%%%%%%%%%%%%%%%%%%%%%%%%%%%%%%%%%%%%%%%%%%%%%%%%%%%%%%%%%%%%%%%
\section{One-loop calculations in QED \label{sec:oneloop}}

In the following calculations, we use the $\overline{\rm MS}$ scheme with a spacetime dimension of $d = 4 - \epsilon$ and a renormalization scale of $\mu$, while compensating a mass dimension by a modified renormalization scale $\tilde \mu$ defined by
\eqs{
{\tilde \mu}^{2} = \mu^{2} \frac{e^{\gamma_{E}}}{4 \pi}
}
with $\gamma_{E} \simeq 0.577$ being Euler's constant.
One-loop functions are summarized in \cref{sec:loopfunc}.

\subsection{Scalar \label{sec:scalar}}
The Lagrangian density is given by%
\footnote{
The following procedure is simplified with the hep-th notation since $A_{\mu}$ has a mass dimension $1$ and is not renormalized due to the Ward-Takahashi identity.
In this case, one needs to multiply $e^{2}$ when translating ${\cal M}_{TAA}$ to ${\cal M}_{\rm dec}$ since the gauge field is not canonically normalized.
}
\eqs{
{\cal L} = - \frac{1}{4} F_{\mu \nu}^{2} - \frac{1}{2 \xi} (\partial_{\mu} A^{\mu})^{2} + |D_{\mu} \phi|^{2} - m^{2} |\phi|^{2} - \frac{1}{4} \lambda |\phi|^{4} \,,
}
with $D_{\mu} = \partial_{\mu} - i q e A_{\mu}$ being the gauge covariant derivative for a charge $q$.
We have integrated out the NL and (anti-)ghost fields.
Parameters are a gauge coupling $e$, a scalar mass $m$, a quartic coupling $\lambda$, and a gauge fixing parameter $\xi$.
Multiplicative renormalization is set for fields as $\phi = Z_{2}^{1/2} {\bar \phi}$ and $A_{\mu} = Z_{3}^{1/2} {\bar A}_{\mu}$ and for parameters as $Z_{2} Z_{3}^{1/2} e = Z_{1} {\tilde \mu}^{\epsilon / 2} {\bar e}$, $Z_{2} Z_{3} e^{2} = Z_{4} {\tilde \mu}^{\epsilon} {\bar e}^{2}$ (i.e., $Z_{2} Z_{4}= Z_{1}^{2}$), $Z_{2} m^{2} = Z_{m} {\bar m}^{2}$, $Z^{2} \lambda = Z_{\lambda} {\tilde \mu}^{\epsilon} {\bar \lambda}$, and $Z_{3} / \xi = Z_{5} / {\bar \xi}$.
The Lagrangian density can be written in the form of renormalized perturbation theory as
\eqs{
{\cal L} =& - \frac{1}{4} {\bar F}_{\mu \nu}^{2} - \frac{1}{2 {\bar \xi}} (\partial_{\mu} {\bar A}^{\mu})^{2} + |\partial_{\mu} {\bar \phi}|^{2} - {\bar m}^{2} |{\bar \phi}|^{2} \\
& - \frac{1}{4} Z_{\lambda} {\tilde \mu}^{\epsilon} {\bar \lambda} |{\bar \phi}|^{4} + i q Z_{1} {\tilde \mu}^{\epsilon/2} {\bar e} {\bar A}^{\mu} ({\bar \phi}^{*} \partial_{\mu} {\bar \phi} - \partial_{\mu} {\bar \phi}^{*} {\bar \phi}) + q^{2} Z_{4} {\tilde \mu}^{\epsilon} {\bar e}^{2} {\bar A}_{\mu}^{2} |{\bar \phi}|^{2} \\
& - \frac{1}{4} (Z_{3} - 1) {\bar F}_{\mu \nu}^{2} - \frac{1}{2 {\bar \xi}} (Z_{5} - 1) (\partial_{\mu} {\bar A}^{\mu})^{2} + (Z_{2} - 1) |\partial_{\mu} {\bar \phi}|^{2} - (Z_{m} - 1) {\bar m}^{2} |{\bar \phi}|^{2} \,.
}
The Ward-Takahashi identity warrants that $Z_{1} = Z_{2} = Z_{4}$, $Z_{3}$ is independent of ${\bar \xi}$, and $Z_{5} = 1$.
It follows that
\eqs{
& \beta^{\epsilon}_{e} = - \frac{\bar e}{2} \epsilon \left( 1 - \frac{\bar e}{2} \frac{\partial \ln Z_{3}}{\partial {\bar e}} \right)^{-1} \,, \\
& \beta^{\epsilon}_{\lambda} = - {\bar \lambda} \epsilon \left( 1 - 2 {\bar \lambda} \frac{\partial \ln Z_{2}}{\partial {\bar \lambda}} + {\bar \lambda} \frac{\partial \ln Z_{\lambda}}{\partial {\bar \lambda}} \right)^{-1} \,, \\
& \beta_{m} = \frac{\bar m}{2} \beta^{\epsilon}_{e} \left( \frac{\partial \ln Z_{2}}{\partial {\bar e}} - \frac{\partial \ln Z_{m}}{\partial {\bar e}} \right) + \frac{\bar m}{2} \beta^{\epsilon}_{\lambda} \left( \frac{\partial \ln Z_{2}}{\partial {\bar \lambda}} - \frac{\partial \ln Z_{m}}{\partial {\bar \lambda}} \right) + \frac{\bar m}{2} \beta_{\xi} \left( \frac{\partial \ln Z_{2}}{\partial {\bar \xi}} - \frac{\partial \ln Z_{m}}{\partial {\bar \xi}} \right) \,, \\
& \beta_{\xi} = - {\bar \xi} \beta^{\epsilon}_{e} \frac{\partial \ln Z_{3}}{\partial {\bar e}} - {\bar \xi} \beta^{\epsilon}_{\lambda} \frac{\partial \ln Z_{2}}{\partial {\bar \lambda}} \,.
}

$Z_{3} - 1$ and $Z_{5} - 1$ can be determined via loop corrections to the two point correlation function of the gauge boson:
\eqs{
i {\bar \Pi}^{\mu \nu} = i \Pi^{\mu \nu} - i (Z_{3} - 1) (k^{2} g^{\mu \nu} - k^{\mu} k^{\nu}) - i \frac{1}{\bar \xi} (Z_{5} - 1) k^{\mu} k^{\nu} \,,
}
where $k$ denotes the gauge boson momentum.
The one-loop vacuum polarization is given by
\eqs{
i \Pi^{\mu \nu} &= (i q {\bar e})^{2} i^{2} {\tilde \mu}^{\epsilon} \int \frac{d^{d} \ell}{(2 \pi)^{d}} \frac{(2 \ell + k)^{\mu} (2 \ell + k)^{\nu}}{[\ell^{2} - {\bar m}^{2}] [(\ell+k)^{2} - {\bar m}^{2}]}
+ (2 i q^{2} {\bar e}^{2} g^{\mu \nu}) i  {\tilde \mu}^{\epsilon} \int \frac{d^{D} \ell}{(2 \pi)^{D}} \frac{1}{[\ell^{2} - {\bar m}^{2}]} \\
& = \frac{i q^{2} {\bar e}^{2}}{16 \pi^{2}} \left( \left[ 4 B_{22} - 2 A \right] g^{\mu \nu} + \left[ 4 B_{21} + 4 B_{1} + B_{0} \right] k^{\mu} k^{\nu} \right) \,.
}
Noting that
\eqs{
4 B_{21} + 4 B_{1} + B_{0} &= \frac{4}{3 k^{2}} \left[ A - {\bar m}^{2} B_{0} + \frac{k^{2}}{4} B_{0} - {\bar m}^{2} + \frac{k^{2}}{6} \right] \\
&= - \frac{1}{k^{2}} [4 B_{22} - 2 A] \,,
}
which ensures the Ward-Takahashi identity,
one finds $i \Pi^{\mu \nu} = (k^{2} g^{\mu \nu} - k^{\mu} k^{\nu}) i \Pi$ and
\eqs{
i \Pi = i \frac{4}{3 k^{2}} \frac{q^{2} {\bar e}^{2}}{16 \pi^{2}} \left[ - A + {\bar m}^{2} B_{0} - \frac{k^{2}}{4} B_{0} + {\bar m}^{2} - \frac{k^{2}}{6} \right] \,.
}
The pole is canceled with
\eqs{
\label{eq:scalarZ3}
Z_{3} - 1 = - \frac{2}{3} \frac{q^{2} {\bar e}^{2}}{16 \pi^{2}} \frac{1}{\epsilon}
}
and $Z_{5} - 1 = 0$.
Thus the four-dimension $\beta$ function is given by
\eqs{
\label{eq:scalarbeta}
\beta_{e} = - \frac{{\bar e}^{2}}{4} \frac{\partial \left( \ln Z_{3} \right)^{\rm residue}}{\partial {\bar e}} = \frac{1}{3} \frac{q^{2} {\bar e}^{3}}{16 \pi^{2}} \,.
}

The contribution from the one-loop diagram with the scalar mass term and $\eta$ term inserted is given by
\eqs{
{\cal M}_{|\phi|^{2}} &= 2 ( {\bar m^{2}} - {\bar \eta} k^{2} ) \left( (i q {\bar e})^{2} i^{3} \int \frac{d^{4} \ell}{(2 \pi)^{4}} \frac{(2 \ell + k_{1}) \cdot \epsilon^{*}_{1} \, (2 \ell + 2 k_{1} + k_{2}) \cdot \epsilon^{*}_{2}}{[\ell^{2} - {\bar m}^{2}] [(\ell + k_{1})^{2} - {\bar m}^{2}] [(\ell + k_{1} + k_{2})^{2} - {\bar m}^{2}]} + [1 \leftrightarrow 2]  \right. \\
& \left. + 2 i q^{2} {\bar e}^{2} \epsilon^{*}_{1} \cdot \epsilon^{*}_{2} \, i^{2} \int \frac{d^{4} \ell}{(2 \pi)^{4}} \frac{1}{[\ell^{2} - {\bar m}^{2}] [(\ell + k_{1} + k_{2})^{2} - {\bar m}^{2}] } \right) \\
&= 2 ( {\bar m^{2}} - {\bar \eta} k^{2} ) \frac{q^{2} {\bar e}^{2}}{16 \pi^{2}} \left( \left[ - 4 C_{24} + B_{0}(k^{2}) \right]\epsilon^{*}_{1} \cdot \epsilon^{*}_{2} + \left[ - 4 C_{23} - 4 C_{12} \right]
 k_{2} \cdot \epsilon^{*}_{1} \, k_{1} \cdot \epsilon^{*}_{2} + [1 \leftrightarrow 2] \right) \,.
}
Noting that
\eqs{
- 4 C_{23} - 4 C_{12} &= \frac{1}{2 k^{2}} \left[ 2 m^{2} C_{0} + 1 \right] \\
&= - \frac{1}{2 k^{2}} \left[ - 4 C_{24} + B_{0} (k^{2}) \right] \,,
}
one finds
\eqs{
{\cal M}_{|\phi|^{2}} = - 4 \frac{{\bar m}^{2} - {\bar \eta} k^{2} }{k^{2}} \frac{q^{2} {\bar e}^{2}}{16 \pi^{2}} \left[ 2 {\bar m}^{2} C_{0} + 1 + [1 \leftrightarrow 2] \right] \left( k_{1} \cdot k_{2} \, \epsilon^{*}_{1} \cdot \epsilon^{*}_{2} - k_{2} \cdot \epsilon^{*}_{1} \, k_{1} \cdot \epsilon^{*}_{2} \right) \,.
}
With
\eqs{
2 {\bar m}^{2} C_{0} + 1 = - \frac{k^{2}}{12 {\bar m}^{2}} I_{s} \left( \frac{k^{2}}{{\bar m}^{2}} \right) \,,
}
the matrix element is
\eqs{
{\cal M}_{|\phi|^{2}} = \frac{2}{3} \frac{q^{2} {\bar e}^{2}}{16 \pi^{2}} \frac{{\bar m}^{2} - {\bar \eta} k^{2} }{{\bar m}^{2}} I_{s} \left( \frac{k^{2}}{{\bar m}^{2}} \right) \left( k_{1} \cdot k_{2} \, \epsilon^{*}_{1} \cdot \epsilon^{*}_{2} - k_{2} \cdot \epsilon^{*}_{1} \, k_{1} \cdot \epsilon^{*}_{2} \right) \,.
}

Let us see how we obtain the above result in Pauli–Villars regularization.
Above the Pauli-Villars mass scale, $\beta_{e} = 0$ and the trace of energy-momentum tensor is replaced by
\eqs{
T^{\mu}_{~ \mu} \supset 2 {\bar m}^{2} |{\bar \phi}|^{2} + 2 {\bar \eta} \partial^{2} |{\bar \phi}|^{2} + 2 {\bar m}_{\rm PV}^{2} |{\bar \phi}_{\rm PV}|^{2} + 2 {\bar \eta}_{\rm PV} \partial^{2} |{\bar \phi}_{\rm PV}|^{2} \,,
}
where ${\bar \phi}_{\rm PV}$ is a Pauli-Villars partner with a wrong statistics.
As a result, the matrix element is replaced by
\eqs{
{\cal M}_{TAA} = - \frac{2}{3} \frac{q^{2} {\bar e}^{2}}{16 \pi^{2}} \left( \frac{{\bar m}_{\rm PV}^{2} - {\bar \eta}_{\rm PV} k^{2} }{{\bar m}_{\rm PV}^{2}} I_{s} \left( \frac{k^{2}}{{\bar m}_{\rm PV}^{2}} \right) - \frac{{\bar m}^{2} - {\bar \eta} k^{2} }{{\bar m}^{2}} I_{s} \left( \frac{k^{2}}{{\bar m}^{2}} \right) \right) \left( k_{1} \cdot k_{2} \, \epsilon^{*}_{1} \cdot \epsilon^{*}_{2} - k_{1} \cdot \epsilon^{*}_{2} \, k_{1} \cdot \epsilon^{*}_{2} \right) \,.
}
After integrating out the Pauli-Villars partner, i.e., ${\bar m}_{\rm PV}^{2} \to \infty$, one reproduces the above result.

\subsection{Fermion \label{sec:fermion}}
The Lagrangian density is given by
\eqs{
{\cal L} = - \frac{1}{4} F_{\mu \nu}^{2 } - \frac{1}{2 \xi} (\partial_{\mu} A^{\mu})^{2} - \frac{1}{2} i D_{\mu}\overline{\psi} \gamma^{\mu} \psi + \frac{1}{2} \overline{\psi} \gamma^{\mu} i D_{\mu} \psi - m \overline{\psi} \psi \,,
}
with $D_{\mu} = \partial_{\mu} - i q e A_{\mu}$ being the gauge covariant derivative for a charge $q$.
We have integrated out the Nakanishi-Lautrup and (anti-)ghost fields.
Parameters are a gauge coupling $e$, a fermion mass $m$, and a gauge fixing parameter $\xi$.

Multiplicative renormalization is set for fields as $\psi = Z_{2}^{1/2} {\bar \psi}$%
\footnote{
Note that a bar for a renormalized quantity is different from a overline for a Dirac bar.
} 
and $A_{\mu} = Z_{3}^{1/2} {\bar A}_{\mu}$ and for parameters as $Z_{2} Z_{3}^{1/2} e = Z_{1} {\tilde \mu}^{\epsilon / 2} {\bar e}$, $Z_{2} m = Z_{m} {\bar m}$, and $Z_{3} / \xi = Z_{4} / {\bar \xi}$.
The Lagrangian density can be written in the form of renormalized perturbation theory as
\eqs{
{\cal L} =& - \frac{1}{4} {\bar F}_{\mu \nu}^{2} - \frac{1}{2 {\bar \xi}} (\partial_{\mu} {\bar A}^{\mu})^{2} - \frac{1}{2} i D_{\mu} \overline{\bar \psi} \gamma^{\mu} {\bar \psi} + \frac{1}{2} \overline{\bar \psi} \gamma^{\mu} i D_{\mu} {\bar \psi} - {\bar m} \overline{\bar \psi} {\bar \psi} + q Z_{1} {\bar e} {\tilde \mu}^{\epsilon / 2} {\bar A}_{\mu} \overline{\bar \psi} \gamma^{\mu} {\bar \psi} \\
& - \frac{1}{4} (Z_{3} - 1) {\bar F}_{\mu \nu}^{2} - \frac{1}{2 {\bar \xi}} (Z_{4} - 1) (\partial_{\mu} {\bar A}^{\mu})^{2} - (Z_{2} - 1) \frac{1}{2} i D_{\mu} \overline{\bar \psi} \gamma^{\mu} {\bar \psi} + (Z_{2} - 1) \frac{1}{2} \overline{\bar \psi} \gamma^{\mu} i D_{\mu} {\bar \psi} - (Z_{m} - 1) {\bar m} \overline{\bar \psi} {\bar \psi} \,.
}
The Ward-Takahashi identity warrants that $Z_{1} = Z_{2}$, $Z_{3}$ is independent of ${\bar \xi}$, and $Z_{4} = 1$.
It follows that
\eqs{
& \beta^{\epsilon}_{e} = - \frac{\bar e}{2} \epsilon \left( 1 - \frac{\bar e}{2} \frac{\partial \ln Z_{3}}{\partial {\bar e}} \right)^{-1} \,, \\
& \beta_{m} = {\bar m} \beta^{\epsilon}_{e} \left( \frac{\partial \ln Z_{2}}{\partial {\bar e}} - \frac{\partial \ln Z_{m}}{\partial {\bar e}} \right) + {\bar m} \beta_{\xi} \left( \frac{\partial \ln Z_{2}}{\partial {\bar \xi}} - \frac{\partial \ln Z_{m}}{\partial {\bar \xi}} \right) \,, \\
& \beta_{\xi} = - \beta^{\epsilon}_{e} \frac{\partial \ln Z_{3}}{\partial {\bar e}}{\bar \xi} \,.
}

$Z_{3} - 1$ and $Z_{4} - 1$ can be determined via loop corrections to the two point correlation function of the gauge boson:
\eqs{
i {\bar \Pi}^{\mu \nu} = i \Pi^{\mu \nu} - i (Z_{3} - 1) (k^{2} g^{\mu \nu} - k^{\mu} k^{\nu}) - i \frac{1}{\bar \xi} (Z_{4} - 1) k^{\mu} k^{\nu} \,,
}
where $k$ denotes the gauge boson momentum.
The one-loop vacuum polarization is given by
\eqs{
i \Pi^{\mu \nu} &= (i q {\bar e})^{2} (-1) i^{2} {\tilde \mu}^{\epsilon} \int \frac{d^{d} \ell}{(2 \pi)^{d}} \frac{{\rm tr} \left[ \gamma^{\mu} ({\slashed \ell} + {\slashed k} + {\bar m}) \gamma^{\nu} ({\slashed \ell} + {\bar m}) \right]}{[\ell^{2} - {\bar m}^{2}] [(\ell+k)^{2} - {\bar m}^{2}]} \\ 
& = - \frac{i q^{2}}{16 \pi^{2}} 4 \left( \left[(-2 + \epsilon) B_{22} - k^{2} \left( B_{21} + B_{1} \right) + {\bar m}^{2} B_{0} \right] g^{\mu \nu} + \left[ 2 B_{21} + 2 B_{1} \right] k^{\mu} k^{\nu} \right) \,.
}
Noting that
\eqs{
2 B_{21} + 2 B_{1} &= \frac{2}{3 k^{2}} \left[ A - {\bar m}^{2} B_{0}- \frac{k^{2}}{2} B_{0} - {\bar m}^{2} + \frac{k^{2}}{6} \right] \\
&= - \frac{1}{k^{2}} [(-2 + \epsilon) B_{22} - k^{2} \left( B_{21} + B_{1} \right) + {\bar m}^{2} B_{0}] \,,
}
which ensures the Ward-Takahashi identity,
one finds $i \Pi^{\mu \nu} = (k^{2} g^{\mu \nu} - k^{\mu} k^{\nu}) i \Pi$ and
\eqs{
i \Pi = i \frac{8}{3 k^{2}} \frac{q^{2} {\bar e}^{2}}{16 \pi^{2}} \left[ A - m^{2} B_{0}- \frac{k^{2}}{2} B_{0} - m^{2} + \frac{k^{2}}{6} \right] \,.
}
The pole is canceled with
\eqs{
\label{eq:fermionZ3}
Z_{3} - 1 = - \frac{8}{3} \frac{q^{2} {\bar e}^{2}}{16 \pi^{2}} \frac{1}{\epsilon}
}
and $Z_{4} - 1 = 0$.
Thus the four-dimension $\beta$ function is given by
\eqs{
\label{eq:fermionbeta}
\beta_{e} = - \frac{{\bar e}^{2}}{4} \frac{\partial \left( \ln Z_{3} \right)^{\rm residue}}{\partial {\bar e}} = \frac{4}{3} \frac{q^{2} {\bar e}^{3}}{16 \pi^{2}} \,.
}

$d$-dimension flat-spacetime energy-momentum tensor is given by%
\footnote{
Curved-spacetime energy-momentum tensor takes the same form with $D_{\mu}$ being the gauge, Local Lorentz, and diffeomorphism covariant derivative.
}
\eqs{
T_{\mu \nu} =& - g^{\lambda \kappa} F_{\mu \lambda} F_{\nu \kappa} - \frac{1}{4}  \left( i D_{\mu} \overline{\psi} \gamma_{\nu} + i D_{\nu} \overline{\psi} \gamma_{\mu} \right) \psi + \frac{1}{4} \overline{\psi} \left( i D_{\mu} \gamma_{\nu} + i D_{\nu} \gamma_{\mu} \right) \psi \\
& - g_{\mu \nu} \left( - \frac{1}{4} F_{\mu \nu}^{2 } - \frac{1}{2} i D_{\mu}\overline{\psi} \gamma^{\mu} \psi + \frac{1}{2} \overline{\psi} \gamma^{\mu} i D_{\mu} \psi - m \overline{\psi} \psi \right) \,.
}
Taking a {\it classical} trace, one finds
\eqs{
\label{eq:fermionclassicalT}
(T^{\mu}_{~ \mu})_{\rm class} = - \frac{1}{4} \epsilon F_{\mu \nu}^{2} + m \overline{\psi} \psi - \left(\frac{3}{2} - \frac{\epsilon}{2} \right) ({\rm e.o.m}) \,,
}
where
\eqs{
({\rm e.o.m}) = \left( - i {\slashed D} \overline{\psi}  - m \overline{\psi} \right) \psi + \overline{\psi} \left(i {\slashed D} - m \right) \psi \,.
}
The first term of $( T^{\mu}_{~ \mu} )_{\rm class}$ vanishes at the {\it classical} level as $\epsilon \to 0$, but not at the {\it quantum} level.
This contribution provides $A_{\rm anom}$.
The leading contribution to ${\cal M}_{TAA}$ arises from the following trace of energy-momentum tensor:
\eqs{
T^{\mu}_{~ \mu} \supset \frac{2}{3} \frac{q^{2} {\bar e}^{2}}{16 \pi^{2}} {\bar F}_{\mu \nu}^{2} + {\bar m} \overline{\bar \psi} {\bar \psi} \,.
}
The first term arises from the gauge kinetic term proportional to $\epsilon$ in \cref{eq:fermionclassicalT}.
Its coefficient is obtained from the leading contribution to the wave function renormalization $Z_{3}$ of the gauge field [see \cref{eq:fermionZ3}].
Note that the leading contribution to $Z_{3}$ also determines the leading contribution to the $\beta$ function $\beta_{e}$ [see \cref{eq:fermionbeta}]
The all-order form that is often quoted,
\eqs{
T^{\mu}_{~ \mu} = \frac{\beta_{e}}{2 e} [ F_{\mu \nu}^{2} ] +  ({\bar m} - \beta_{m}) [ \overline{\psi} \psi ] \,,
}
is obtained after renormalization of composite operators~\cite{Adler:1976zt}.

The matrix element is
\eqs{
{\cal M}_{TAA} = {\cal M}_{F^{2}} + {\cal M}_{\overline{\psi} \psi} \,.
}
The first term arises from the tree-level diagram with the gauge kinetic term inserted:
\eqs{
{\cal M}_{F^{2}} = - \frac{8}{3} \frac{q^{2} {\bar e}^{2}}{16 \pi^{2}} \left( k_{1} \cdot k_{2} \, \epsilon^{*}_{1} \cdot \epsilon^{*}_{2} - k_{2} \cdot \epsilon^{*}_{1} \, k_{1} \cdot \epsilon^{*}_{2} \right) \,.
}
The second term is a one-loop contribution from the fermion mass term inserted:
\eqs{
{\cal M}_{\overline{\psi} \psi} =& {\bar m} \left( (i q {\bar e})^{2} (- 1) i^{3} \int \frac{d^{4} \ell}{(2 \pi)^{4}} \frac{{\rm tr} \left[ ({\slashed \ell} + {\slashed k}_{1} + {\slashed k}_{2} + {\bar m}) {\slashed \epsilon}^{*}_{2} ({\slashed \ell} + {\slashed k}_{1} + {\bar m}) {\slashed \epsilon}^{*}_{1} ({\slashed \ell} + {\bar m}) \right]}{[\ell^{2} - {\bar m}^{2}] [(\ell + k_{1})^{2} - {\bar m}^{2}] [(\ell + k_{1} + k_{2})^{2} - {\bar m}^{2}]} + [1 \leftrightarrow 2] \right) \\
=& 4 {\bar m}^{2} \frac{q^{2} {\bar e}^{2}}{16 \pi^{2} {\tilde \mu}^{\epsilon}} \left( \left[ \epsilon C_{24} - k^{2} \, C_{23} - k^{2} \, C_{12} - \frac{k^{2}}{2} \, C_{0} + {\bar m}^{2} C_{0} \right] \epsilon^{*}_{1} \cdot \epsilon^{*}_{2} \right. \\
& \left. + \left[ 4 C_{23} + 4 C_{12} +C_{0} \right] k_{2} \cdot \epsilon^{*}_{1} \, k_{1} \cdot \epsilon^{*}_{2} + [1 \leftrightarrow 2] \right) \,.
}
Noting that
\eqs{
4 C_{23} + 4 C_{12} +C_{0} &= - \frac{1}{2 k^{2}} \left[ 2 {\bar m}^{2} C_{0} + 1 - \frac{k^{2}}{2} \, C_{0} \right] \\
&= - \frac{1}{2 k^{2}} \left[ \epsilon C_{24} - k^{2} \, C_{23} - k^{2} \, C_{12} - \frac{k^{2}}{2} \, C_{0} + {\bar m}^{2} C_{0} \right] \,,
}
one finds
\eqs{
{\cal M}_{\overline{\psi} \psi} = 8 \frac{{\bar m}^{2}}{k^{2}} \frac{q^{2} {\bar e}^{2}}{16 \pi^{2}} \left[ 2 {\bar m}^{2} C_{0} + 1 - \frac{k^{2}}{2} C_{0} + [1 \leftrightarrow 2] \right] \left( k_{1} \cdot k_{2} \, \epsilon^{*}_{1} \cdot \epsilon^{*}_{2} - k_{2} \cdot \epsilon^{*}_{1} \, k_{1} \cdot \epsilon^{*}_{2} \right) \,.
}
With
\eqs{
2 {\bar m}^{2} C_{0} + 1 - \frac{k^{2}}{2} C_{0} = \frac{k^{2}}{6 {\bar m}^{2}} I_{f} \left( \frac{k^{2}}{m^{2}} \right) \,,
}
the matrix element is
\eqs{
{\cal M}_{\overline{\psi} \psi} = \frac{8}{3} \frac{q^{2} {\bar e}^{2}}{16 \pi^{2}} I_{f} \left( \frac{k^{2}}{m^{2}} \right) \left( k_{1} \cdot k_{2} \, \epsilon^{*}_{1} \cdot \epsilon^{*}_{2} - k_{2} \cdot \epsilon^{*}_{1} \, k_{1} \cdot \epsilon^{*}_{2} \right) \,.
}
Collecting the two contributions, one obtains
\eqs{
{\cal M}_{TAA} = \frac{8}{3} \frac{q^{2} {\bar e}^{2}}{16 \pi^{2}} \left( 1 - I_{f} \left( \frac{k^{2}}{{\bar m}^{2}} \right) \right) \left( k_{1} \cdot k_{2} \, \epsilon^{*}_{1} \cdot \epsilon^{*}_{2} - k_{2} \cdot \epsilon^{*}_{1} \, k_{1} \cdot \epsilon^{*}_{2} \right) \,.
}
We remark that $I_{f} (0) = 1$ and thus a heavy (${\bar m}^2 \gg k^{2}$) fermion does not contribute to ${\cal M}_{TAA}$.
Meanwhile, $I_{f} (\infty) = 0$ and thus a light (${\bar m}^2 \ll k^{2}$) fermion indeed contributes to ${\cal M}_{TAA}$.

Let us see how we obtain the above result in Pauli–Villars regularization.
Above the Pauli-Villars mass scale, $\beta_{e} = 0$ and the trace of energy-momentum tensor is replaced by
\eqs{
T^{\mu}_{~ \mu} \supset {\bar m} \overline{\bar \psi} {\bar \psi} + {\bar m}_{\rm PV} \overline{\bar \psi}_{\rm PV} {\bar \psi}_{\rm PV} \,,
}
where ${\bar \psi}_{\rm PV}$ is a Pauli-Villars partner with a wrong statistics.
As a result, the matrix element is replaced by
\eqs{
{\cal M}_{TAA} = \frac{8}{3} \frac{q^{2} {\bar e}^{2}}{16 \pi^{2}} \left( I_{f} \left( \frac{k^{2}}{{\bar m}_{\rm PV}^{2}} \right) - I_{f} \left( \frac{k^{2}}{{\bar m}^{2}} \right) \right) \left( k_{1} \cdot k_{2} \, \epsilon^{*}_{1} \cdot \epsilon^{*}_{2} - k_{2} \cdot \epsilon^{*}_{1} \, k_{1} \cdot \epsilon^{*}_{2} \right) \,.
}
After integrating out the Pauli-Villars partner, i.e., ${\bar m}_{\rm PV}^{2} \to \infty$, one reproduces the above result.

\subsection{Summary of one-loop functions \label{sec:loopfunc}}
One-loop functions are based on Refs.~\cite{Passarino:1978jh, tHooft:1978jhc} (see also Appendix F of Ref.~\cite{Logan:1999if}).
One point integral is defined as
\eqs{
{\tilde \mu}^{\epsilon} \int \frac{d^{d} \ell}{(2 \pi)^{d}} \frac{1}{\ell^{2} - m^{2}} = \frac{i}{16 \pi^{2}} A (m^{2}) \,.
}
The explicit form is
\eqs{
A (m^{2}) = m^{2} \left( \frac{2}{\epsilon} - \ln \left( \frac{m^{2}}{\mu^{2}} \right) + 1 \right) \,.
}

Two point integrals are defined as
\eqs{
{\tilde \mu}^{\epsilon} \int \frac{d^{d} \ell}{(2 \pi)^{d}} \frac{1; \ell_{\mu}; \ell_{\mu} \ell_{\nu}}{[\ell^{2} - m_{1}^{2}] [(\ell+k)^{2} - m_{2}^{2}]} = \frac{i}{16 \pi^{2}} B_{0; \mu; \mu \nu} (k^{2}; m_{1}^{2}, m_{2}^{2}) \,,
}
where
\eqs{
& B_{\mu} = k_{\mu} B_{1} \,, \\
& B_{\mu \nu} = g_{\mu \nu} B_{22} + k_{\mu} k_{\nu} B_{21} \,.
}
For our purpose, we can take $m_{1} = m_{2} = m$:
\eqs{
& B_{1} = - \frac{1}{2} B_{0} \,, \\
& B_{22} = \frac{1}{6} \left[ A + 2 {\bar m}^{2} B_{0} - \frac{k^{2}}{2} B_{0} + 2 {\bar m}^{2} - \frac{k^{2}}{3} \right] \,, \\
& B_{21} = \frac{1}{3 k^{2}} \left[ A - {\bar m}^{2} B_{0} + k^{2} B_{0} - {\bar m}^{2} + \frac{k^{2}}{6} \right] \,.
}
The explicit form with a Feynman parameter integral is
\eqs{
B_{0} = \frac{2}{\epsilon} - \int^{1}_{0} dx \ln \left( \frac{m^{2} - x (1 - x) k^{2} - i \epsilon_{\rm ad}}{\mu^{2}} \right) \,.
}

Three point integrals are defined as
\eqs{
{\tilde \mu}^{\epsilon} \int \frac{d^{d} \ell}{(2 \pi)^{d}} \frac{1; \ell_{\mu}; \ell_{\mu} \ell_{\nu}}{[\ell^{2} - m_{1}^{2}] [(\ell+k_{1})^{2} - m_{2}^{2}] [(\ell+k_{1}+k_{2})^{2} - m_{3}^{2}]} = \frac{i}{16 \pi^{2}} C_{0; \mu; \mu \nu} (k_{1}^{2}, k_{2}^{2}, k^{2}; m_{1}^{2}, m_{2}^{2}, m_{3}^{2}) \,,
}
where $k + k_{1} + k_{2} = 0$ and 
\eqs{
& C_{\mu} = k_{1 \mu} C_{11} + k_{2 \mu} C_{12} \,, \\
& C_{\mu \nu} = g_{\mu \nu} C_{24} + k_{1 \mu} k_{1 \nu} C_{21} + k_{2 \mu} k_{2 \nu} C_{22} + \left( k_{1 \mu} k_{2 \nu} +  k_{2 \mu} k_{1 \nu} \right) C_{23} \,.
}
For our purpose, again we can take $m_{1} = m_{2} = m_{3} = m$:
\eqs{
& C_{11} = \frac{1}{k^{2}} \left[ B_{0} (k_{1}^{2}) - B_{0} (k^{2}) - k^{2} \, C_{0} \right] \,, \\
& C_{12} = \frac{1}{k^{2}} \left[ B_{0} (k^{2}) - B_{0} (k_{2}^{2}) \right] \,, \\
& C_{24} = \frac{1}{4} \left[ B_{0} (k^{2}) + 2 {\bar m}^{2} C_{0} + 1 \right] \,, \\
& C_{21} = - \frac{1}{2 k^{2}} \left[ 3 B_{0} (k^{2}) - 3 B_{0} (k^{2}) - 2 k^{2} \, C_{0} \right] \,, \\
& C_{23} = - \frac{1}{2 k^{2}} \left[ 2 B_{0} (k^{2}) - 2 B_{0} (k_{2}^{2})  + 2 {\bar m}^{2} C_{0} + 1 \right] \,, \\
& C_{22} = - \frac{1}{2 k^{2}} \left[ B_{0} (k^{2}) - B_{0} (k_{2}^{2}) \right] \,.
}
The explicit form with Feynman parameter integrals is
\eqs{
C_{0} = - \int_{0}^{1} dx \int_{0}^{1 - x} dy \, \frac{1}{- k^{2} \, x y + m^{2} - i \epsilon_{\rm ad}} \,.
}

\end{document}